\documentclass[preprint2]{aastex}
\usepackage{natbib,latexsym,amsmath}
\usepackage{graphicx}
\usepackage{lscape}
\usepackage{multirow}
\usepackage{longtable}
\usepackage{alltt}
\usepackage{epsfig}
\usepackage{subfig}

\begin{document}

\title{Ca~II and Na~I Quasar Absorption-Line Systems in an Emission-Selected Sample of SDSS DR7 Galaxy/Quasar Projections: I. Sample Selection}

\author{B. Cherinka and R.E. Schulte-Ladbeck}
\affil{Department of Physics \& Astronomy, University of Pittsburgh,
    Pittsburgh,PA 15260}

\email{bac29@pitt.edu}

\begin{abstract}
The aim of this project is to identify low-redshift host galaxies of quasar absorption-line systems by selecting galaxies which are seen in projection onto quasar sightlines.  To this end, we use the Seventh Data Release of the Sloan Digital Sky Survey (SDSS-DR7) to construct a parent sample of 97489 galaxy/quasar projections at impact parameters of up to 100~kpc to the foreground galaxy.  We then search the quasar spectra for absorption line systems of Ca~II and Na~I within $\pm$ 500~km~s$^{-1}$ of the galaxy's velocity.  This yields 92 Ca~II and 16 Na~I absorption systems.  We find that most of the Ca~II and Na~I systems are sightlines through the Galactic disk, through High Velocity Cloud complexes in our halo,  or Virgo cluster sightlines.  Placing constraints on the absorption line rest equivalent width significance ($\ge$3.0$\sigma$), the Local Standard of Rest velocity along the sightline ($\ge$345~km~s$^{-1}$), and the ratio of the impact parameter to the galaxy optical radius ($\le$5.0), we identify 4 absorption line systems that are associated with low-redshift galaxies at high confidence, consisting of two Ca~II systems (one of which also shows Na~I), and two Na~I systems.  These 4 systems arise in blue, $\sim$L$_r^*$ galaxies.  Tables of the 108 absorption systems are provided to facilitate future follow up.  
\end{abstract}

\keywords{catalog, quasar absorption lines, galaxies, Ca~II, Na~I}

\section{Motivation}
The project we describe in this paper was motivated by 
the fact that the Sloan Digital Sky Survey (SDSS) database contains spectroscopically observed galaxy/quasar projections
which can be used to study quasar absorptions-line systems arising from known galaxies with redshifts between 0$<$z$<$0.6.
QSO absorption lines provide a wealth of information on the physical conditions of gaseous structures along random sightlines through the Universe.  
The strongest absorption-line systems seen are the Hydrogen absorption systems 
(Damped Ly$\alpha$ Absorbers or DLAs, sub-DLAs, Lyman-limit, and Ly$\alpha$-forest systems). 
Quasar spectra also exhibit metal-line absorption systems, with Ca~II and Na~I causing the strongest lines at rest-frame optical wavelengths. 
There is much current interest in studying the connection of quasar absorption-line systems with the galaxies or galaxy environments in which they are thought
to arise, since this can give new information on the evolving morphological structures that they represent \citep{will05,wolf05,baugh06,benson10,dekel06,keres05,keres09,steidel10,kimm11,stewart11}.   With this project we seek to discover, and to characterize, new low redshift absorber galaxies with emphasis on Ca~II and Na~I absorbers.

\subsection{Absorber-Galaxy Associations}
There are two main approaches in uncovering the connection between the absorber host galaxy and the absorbing 
gas seen in quasar spectra. The first, more traditional, approach involves selecting quasar absorption-line systems through spectroscopic observations 
of quasars.  This method is independent of the absorber host galaxy, as the line is found first, 
with follow-up imaging to locate the host galaxy. However, this approach has not produced a large body 
of galaxy/quasar pairs. With few galaxy identifications, then for most absorption 
systems the analysis of the host properties relies on what can be gleaned from the study of the absorption 
lines themselves, e.g., \citet{wolf05}.

The second approach in uncovering the connection between the absorbing
gas and the absorber host galaxy is by searching for known
low-redshift galaxies with quasars projected along lines of sights
towards the galaxy.  One well known example of this approach is the dwarf galaxy SBS~1543+593 which gives rise to a DLA.  The
galaxy/quasar pair was first discovered by \citet{reimers98} and then
extensively studied in absorption and emission \citep{schulte04, schulte05, bowen05, rosen06}. 

A few previous attempts have been made at generating large
galaxy/quasar pair catalogs.  \cite{burb90} compiled a catalog of over
400 galaxy/quasar pairs with angular separations $<$10'.
\cite{bukhma00} compiled a catalog from the literature of 8382
galaxy/quasar pairs separated by $<$150 kpc at the galaxy redshift. 

Even more useful than single pairs is the identification of multiple quasars projected behind one particular foreground galaxy,
allowing comparisons of the absorbing gas along multiple
lines-of-sight through the galaxy.  This approach requires the
foreground galaxy to have a large angular extent so its cross section
covers a large sky area for random sightlines and constrains the foreground galaxy to be of extremely
low redshift. \cite{crampton97} identified 16 AGN behind the Magellanic Clouds
selected via their X-ray emission and 146 quasar candidates behind
nearby galaxies.  \cite{kozlow09} identified 5000 possible quasar
candidates behind the Magellanic Clouds using a mid-IR color selection
scheme. These and other catalogs have provided valuable source lists for follow-up observations of 
galaxy/quasar projections.

\subsection{Project Description}
With the vast numbers of galaxies and quasars found in SDSS, the SDSS is the ideal place to search for
galaxy/quasar projections on a large scale. The power of using the SDSS for this project is that it provides spectra. 
The SDSS spectra cover the wavelength range from 3900~\AA{} to 9200~\AA{}. The pixel size is 69~km~s$^{-1}$ (varying from 0.9~\AA{} at the
blue end, to 2.1~\AA{} at the red end of the spectra); this gives a resolution of about 170~km~s$^{-1}$. Additional information
about the seventh and final SDSS data release (SDSS-DR7) can be obtained from \citet{sdssdr7}

The first goal of this project is to identify and characterize a large number of galaxy/quasar projections in which
the galaxies are projected onto quasar sightlines.  Section 2 of this paper describes how we generated a catalog of spectroscopic galaxy/quasar pairs, hereafter referred to as the
parent sample, from the SDSS-DR7. We include a discussion of the basic characteristics of the parent sample.  

The second goal of this project is to find Ca~II and Na~I absorbers associated with the galaxies in
the galaxy/quasar parent sample. The lines that produce the
strongest interstellar absorption lines at optical wavelengths are the Ca~II doublet at $\lambda\lambda$~3934.77, 3969.59~\AA{}~(K, H) and the Na~I~D doublet at
$\lambda\lambda$~5891.58, 5897.56~\AA{}~(D2, D1).  Both Ca~II and Na~I have ionization potentials that are smaller 
than that of H~I, 11.9~eV and 5.1~eV versus 13.6~eV, respectively, and
are therefore thought to be probes of the neutral gas in and around galaxies. In section 3, we describe the identification and selection of the Ca~II and Na~I absorber galaxies; in section 4, we analyze and discuss the basic observational characteristics of the Ca~II and Na~I absorbers, and in section 5, we describe the absorber systems attributed to extragalactic sources.

In section 6 we summarize the characteristics of the new absorbers and their host galaxies and make suggestions for follow-up studies.

\section{Galaxy-Quasar Parent Sample}
The SDSS-DR7 quasar catalog provides the parent quasar sample \citep{schneider10}. This catalog of bona fide quasars, 
that have redshifts checked by eye and luminosities 
and line widths that meet the formal quasar definition, contains 105783 spectroscopically confirmed quasars, 
and represents the final product of the 
SDSS-I and SDSS-II quasar survey. The quasar catalog uses a cosmology with H$_o$=70 km~s$^{-1}$, ${\Omega}_m$=0.3, and
${\Omega}_{\Lambda}$=0.7, which we adopted for the remainder of this paper. 

The SDSS-DR7 SpecPhoto View provided the parent galaxy sample.
The SpecPhoto table includes only those objects where the SpecObj is a sciencePrimary, and the BEST PhotoObj is a PRIMARY object and the object has clean spectra. 
We selected from this table all objects that were morphologically typed as a galaxy (type = galaxy) and have the spectrum of a galaxy 
(specClass = galaxy). This returned 798948 galaxy spectra. 

\subsection{Parent Sample Definition}

We identified all pairs for which 
\begin{enumerate}
\item the quasar's spectrum is projected within 100~kpc of a galaxy's spectrum and
\item the quasar's spectrum has a redshift which is larger than the galaxy's, $z_{QSO}>z_{gal}$, by an amount equal to the total redshift error, $\sigma$, 
of the galaxy and the quasar,
 $z_{QSO}-z_{gal}$ $>$ $\sqrt{\sigma_{z_{QSO}}^2+\sigma_{z_{gal}}^2}$.
\end{enumerate}
Steps 1 and 2 yielded 97489 pairs of galaxy/quasar spectra. We refer to these pairs as the galaxy-quasar parent sample.  The full list of pairs is available for download as a FITS table with the online version of this paper, a sample of which is provided in Table~\ref{tab:parent}.  Prior to the absorber selection we also apply an additional constraint on the parent sample where
\begin{enumerate}
\setcounter{enumi}{2}
\item we remove cases in which intervening Ly$\alpha$ forest lines potentially cause misidentification.  
\end{enumerate} 
We perform step 3 for both Ca~II and Na~I, thus creating two parent samples, one of 95651 pairs (Na~I) and one of 81912 pairs (Ca~II).  For the remainder of this section we discuss the overall parent sample of 97489 pairs.  

An issue that affects the construction of our sample is the 55'' fiber collision problem in SDSS.  On a given tile in SDSS, there are a maximum of 592 possible fibers for placement onto a viable target.  The spacing between each fiber on a tile is 55''.  Targets that lie less than 55'' from each other are spectroscopically unobservable unless recovered with overlapping tiles on the sky.  About 6$\%$ of all SDSS targets are missed in this manner \citep{blant03}.  With approximately 1.1 million objects targeted for spectroscopy, 6$\%$ corresponds to $\sim$66000 objects.  This however does not necessarily imply that the incompleteness of close pairs is $\sim$6$\%$.  The spectroscopic incompleteness of our sample in particular (close pairs within 100~kpc) will be addressed in a future paper.  At the mean redshift for SDSS galaxies (z$\sim$0.1), 55'' corresponds to $\sim$101~kpc.  For the majority of our galaxies, we are searching well within the fiber collision radius.  

\subsection{Parent Sample Characteristics}

\begin{figure}
\centering
\includegraphics[scale=0.9]{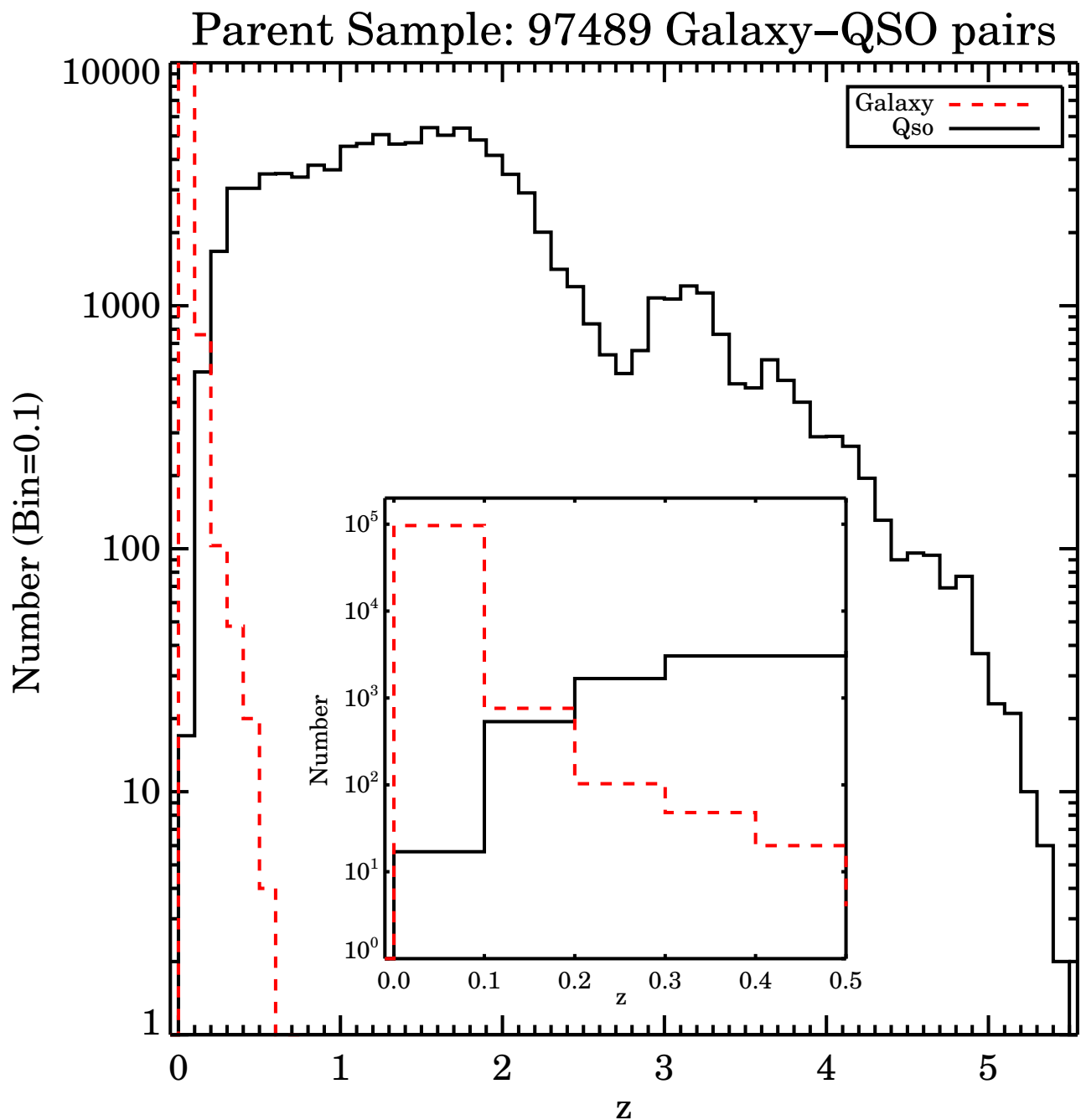}
\caption{Shown here are the redshift distributions for the parent galaxy and quasar sample.  As expected given our catalog constraints, the distribution of quasars does indeed lie behind the distribution of galaxies.  The inset plot shows a zoomed region of z$<$0.5.}
\label{fig:zhisto}
\end{figure}

\begin{figure}
\centering
\includegraphics[scale=0.9]{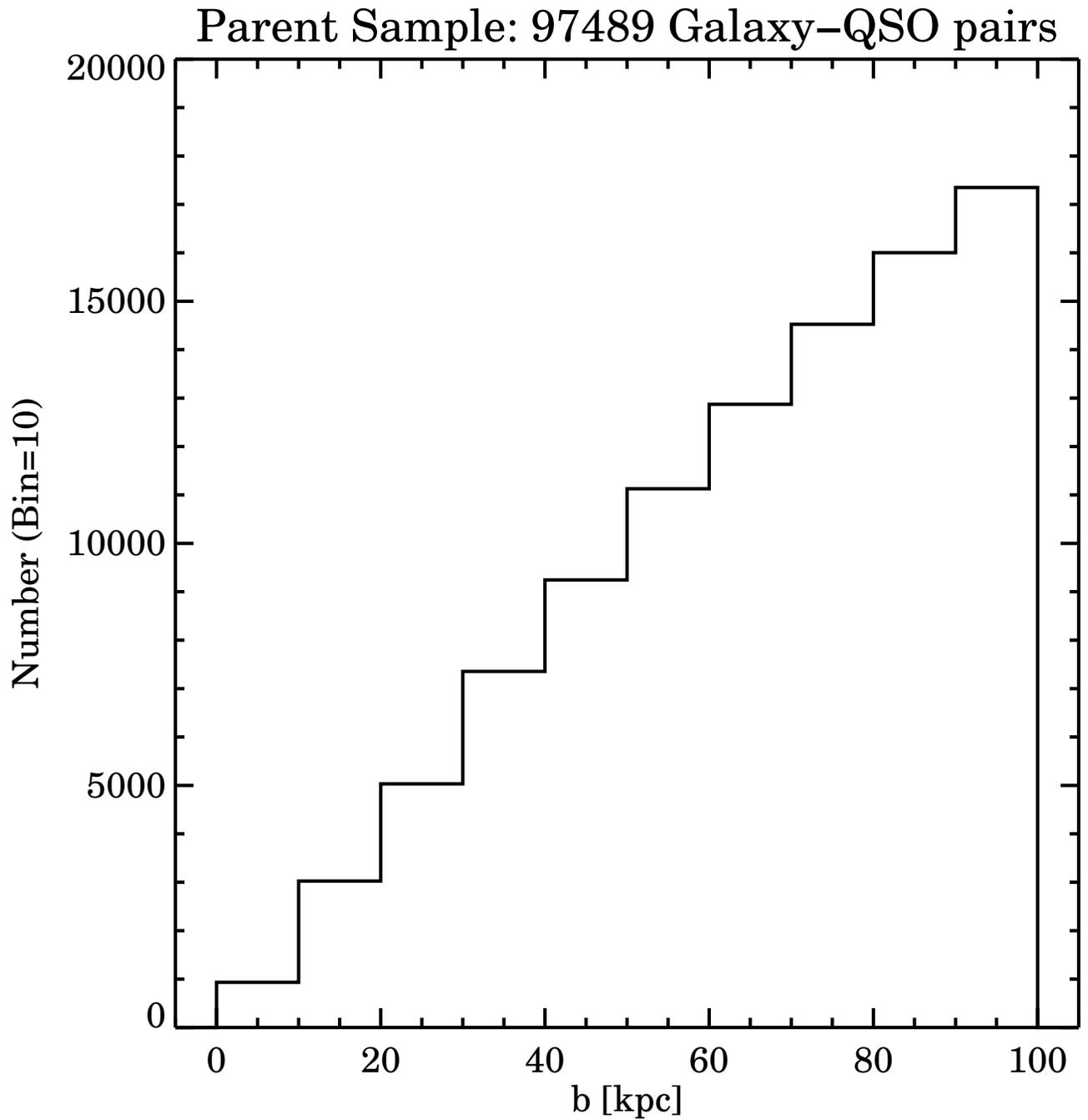}
\caption{Shown here is the distribution of impact parameter in kpc of the parent galaxy-quasar sample.  The large number of pairs seen at large impact parameters is due to the much larger projection of the 100~kpc search radius for low-redshift galaxies, facilitating the detection of many more sightlines.}
\label{fig:bhisto}
\end{figure}

Figure~\ref{fig:zhisto} shows the redshift distributions of the galaxy and quasar spectra in the parent sample of 97489 pairs.  The mean redshift of the galaxy spectra is ($\mu$,$\sigma$)=(0.00649,0.00005), and the mean redshift of the quasar spectra is ($\mu$,$\sigma$)=(1.6,0.8). We see that, indeed, the distribution of QSOs lies mostly behind that of the galaxies.  Figure~\ref{fig:bhisto} displays the impact parameter distribution of the pairs.  We find an increasing number of pairs as the impact parameter increases.  This distribution is expected given the large number of low-redshift galaxies seen in Figure~\ref{fig:zhisto}.  The 100~kpc search radius around these low-redshift galaxies produces a much larger search area on the sky, from which a greater number of sightlines will be observed.  The mean impact parameter of the distribution is ($\mu$,$\sigma$)=(66,24)~kpc.

The spectra which define the galaxy-quasar parent sample do not necessarily belong to unique galaxy/quasar pairs. There are two reasons for this. First, the SDSS database contains multiple spectra for some galaxies. A typical example is a nearby star-forming galaxy for which the SDSS targeted its nucleus plus one or more bright H~II regions. Second, some nearby galaxies cover such a large area on the sky that they intercept two or more quasar sight-lines.  If the coverages of multiple nearby galaxies overlap, then it is also possible to have more than one galaxy intercept the same quasar sightline.

\section{Absorber Galaxies}

We searched all quasar spectra in the parent sample for Ca~II and Na~I absorption lines located at the redshift of the foreground galaxy, permitting a deviation of $\pm$~500~km~s$^{-1}$ from the systemic galaxy velocity to allow for the possibility that the quasar sightline intercepts the galaxy's gas where it has an additional internal motion due to rotation, inflow, or outflow. 

\subsection{Absorber Selection}
Our line identification procedure is as follows. First, we use an automated line finder to create a candidate list of Ca~II and Na~I systems. Second, we review the candidate features by eye to produce final line lists.  We perform the automated line identification on the two reduced parent samples of 95651 and 81912 pairs for Na~I and Ca~II, respectively.   The later visual inspection of the candidate detections also showed a small number of possible intervening Mg~II, Fe~II, or C~IV systems that interfered with a positive identification of the doublet lines. 

We adopted the automated line finder of the Hubble Space Telescope (HST) Quasar Absorption Line Key Project \citep {schneider93} which is optimized to look for weak, unresolved quasar absorption lines. We followed the prescription as outlined in Churchill (in prep.).  The prescription searches for significant unresolved features by weighting each pixel by the Instrumental Spread Function (ISF), a Gaussian that describes how the pixel counts are redistributed below the limiting resolution of the instrument.  The ISF, $\Phi$, is defined as 
\begin{equation}
	\Phi (\lambda^{\prime} - \lambda) = \frac{1}{2\pi\sigma}\exp{[-\frac{(\lambda^{\prime}-\lambda)^2}{2\sigma^2}]} 
\end{equation}
In practice, the ISF is discretized as a Gaussian model, P$_i$, symmetric about a pixel i, that spans M=2J0+1 elements, where J0 = 2p, and p is the number of pixels per resolution element.  The discretized ISF is given as
\begin{align*}
	P_i &= \exp{[-x_{kj}^2]} /  \displaystyle\sum\limits_{i=1}^M \exp{[-x_{kj}^2]} \\
	\tag{2}\\
	x_{kj} &= \frac{\lambda_k - \lambda_j}{\sigma_j^{ISF}} 
\end{align*} 
normalized to insure conservation of counts within the region spanned by the ISF.  
 \addtocounter{equation}{1}
In the above, $\sigma_j^{ISF}$~=~$\lambda_j$~/~(2.35~R) is the Gaussian width of the ISF at pixel j, where R is the spectral resolution, and k~=~j~+~(i-1)~-~J0 is the wavelength index relative to the central pixel j.  

The procedure is outlined as follows: 
\begin{enumerate}      
\item {\bf Weight the Flux Decrement in each Pixel by the ISF:} \\
The weighted equivalent width and uncertainty in pixel j is defined as 
\begin{align}
	ew_j &= -\frac{\Delta \lambda_j}{P^2} \displaystyle\sum\limits_{i=1}^M P_iD_k \\
	\sigma_{ew_j} &= \frac{\Delta \lambda_j}{P^2} \bigg( \displaystyle\sum\limits_{i=1}^M P_i^2 \sigma^2_{D_k} \bigg) ^{1/2}
\end{align}
where
\begin{equation}
	P^2 = \displaystyle\sum\limits_{i=1}^M P_i^2
\end{equation}
D$_k$ =  1-(I$_k$/I$_k^c$) is the flux decrement in pixel k, and $\Delta\lambda_j$ = 0.5$\times$($\lambda_{j+1}$-$\lambda_{j-1}$) is the wavelength dispersion in pixel j.  
\item {\bf Search the Spectrum for Pixels Satisfying the Condition:} \\
\begin{equation}
	\frac{ew_j}{\sigma_{ew_j}} \leq - N_\sigma
\end{equation}
where N$_\sigma$ is a user-specified significance threshold. 
\item {\bf Determine the Pixels Spanned by Feature i:}\\
For a given pixel that satisfies the above condition, identify the start and end pixels of feature i by scanning the spectrum blueward and redward of pixel j until the conditions
\begin{equation}
	\frac{ew_j^-}{\sigma_{ew_j^-}} \geq -1.0 \quad;\quad \frac{ew_j^+}{\sigma_{ew_j^+}} \geq -1.0  
\end{equation}
are met. 
\item { \bf Determine the Significance of the Candidate Feature} \\
Once the pixel span of feature i has been identified, the observed equivalent width, EW, and uncertainty, $\sigma_{EW}$, of feature i are 
\begin{eqnarray}
	EW_i = \displaystyle\sum\limits_{k=j^-}^{j^+} e_k \\
	\sigma^2_{EW_i} = \displaystyle\sum\limits_{k=j^-}^{j^+} \sigma^2_{e_k} 
\end{eqnarray} 
where e$_k$ = $\Delta\lambda_k$D$_k$ is the unweighted equivalent width in pixel k, and $\sigma_e$ = $\Delta\lambda_k$$\sigma_{D_k}$ is the uncertainty in e$_k$.  The final step is to calculate the significance, S, of feature i and check that it satisfies your required significance threshold.   
\begin{equation}
\label{eq:sig}
S_i = \frac{EW_i}{\sigma_{EW_i}} \quad;\quad S_i \geq N_\sigma
\end{equation}
If the candidate feature satisfies the above criteria, then it is kept as a candidate absorption line. 

\end{enumerate}

%Briefly, the prescription consists of first weighting each pixel in the quasar spectrum by the Instrumental Spread Function, a Gaussian that describes the pixel smearing that occurs below the limiting resolution of the instrument.  One then searches this weighted spectrum for pixels that satisfy a specified significance threshold.   Upon locating satisfactory pixels, a candidate line feature is identified and the same significance test is applied to the candidate feature.  If the feature passes the significance threshold, it is kept as a candidate line.  The significance of each pixel is defined as the equivalent width in that pixel over the equivalent width error for that pixel.  The significance of a candidate feature is defined as EW/$\sigma_{EW}$, where EW is the equivalent across the candidate line profile.  We again refer you to Churchill,C. (in prep.) for further details.             

The line finder algorithm expects as input a normalized flux array.  We therefore divided the quasar spectra by the global continuum fit stored in the `spSpec' 1D spectral fits file. In some cases the sought-after doublets were superimposed on a broad emission line in the quasar's spectrum. We thus re-defined a local continuum anchored to the quasar's normalized spectrum within $\pm$~1000~km~s$^{-1}$ of the expected doublets.  

We automatically searched for lines that are located to within $\pm$~500~km~s$^{-1}$ of their expected positions, given the galaxy redshift.  The significance of a candidate line was calculated based on Equation~\ref{eq:sig}.  We chose a significance threshold, N$_{\sigma}$, of two.  All lines with significances less than this threshold were automatically rejected. This step eliminated a very large number of pairs.  Lines were kept as candidates for visual inspection if their significances were equal to or exceeded our specified significance threshold.  

\subsubsection{Ca~II Doublet Sample}
We created three candidate line lists. The first list contained Ca~II doublets. For optically thin gas, the ratio of the primary and secondary line is 2:1, while for saturated lines, the doublet ratio approaches 1:1.  We use this allowed region to define the acceptable strengths of the weaker, Ca~II~H line, i.e. the Ca~II~H equivalent width must be, at least, half as strong as the Ca~II~K line and, at most, equal in strength to the Ca~II~K line.  Questionable candidates, which passed the line finder but not the visual inspection, were moved to a separate list. Among the parent sample for Ca~II systems, we claim 92 detections (A in Table~\ref{tab:parent}), 1641 questionable candidates (Q in Table~\ref{tab:parent}), and 80179 non-detections (N in Table~\ref{tab:parent}).  Main reasons why a doublet was rejected during visual inspection include: (a) possibility of confusion with an interloping non-Ca~II system, (b) a pixel pattern in the doublet region which looked inconsistent with the profile of a real doublet, and (c) the doublet region has a low signal-to-noise.  For the 92 detections, the line finder yields a mean line significance level of the Ca~II~K line of ($\mu$,$\sigma$)=(2.9$\pm$0.1,0.6$\pm$0.1).  The Ca~II absorber catalog consists of 92 detections involving pairs of individual galaxy and quasar spectra.  These 92 detections result from 13 unique galaxy spectra and 61 unique quasar spectra.  Listed in Table~\ref{tab:uniqCaII} are the individual Ca~II absorber galaxies, with the number of quasars found associated with each particular galaxy, along with any notes regarding that particular galaxy.  Table~\ref{tab:uniqCaII} highlights the fact that a low-redshift galaxy may have a large search radius such that many quasar sightlines are intersected, producing a large number of non-unique spectral pairs.  With more than one of these galaxies, it is therefore also possible to have one quasar be considered in multiple pairs with different galaxies.  These facts combine to produce the numbers of unique galaxy and quasar spectra above.  For example, the galaxy SDSS~J21633.70+130153.6 had 21 quasars identified within 100~kpc which exhibited Ca~II absorption located within $\pm$500~km~$s^{-1}$ of the galaxy redshift.  This system would then account for 1 out of 13 unique galaxy spectra, 21 out of 61 unique quasar spectra, and 21 out of 92 Ca~II detections.  Also see the last paragraph of Section 2.2. 

\subsubsection{Na~I Doublet Sample}
The second line list contained Na~I systems. Some of the Na~I doublets are resolved and some are blended.  For partially blended doublets, occasionally the line finder failed to distinguish both components of the doublet, instead treating the entire doublet as the Na~I~D2 line.  It then attempted to find a `false' weaker component for Na~I.  This led us to treat the Na~I candidates differently from the Ca~II candidates.  We required that the line finder identify the `Na~I~D2 line' of the doublet only, then visually rejected as questionable candidates lines showing profile shapes that appear inconsistent with a blended physical doublet.  As with the Ca~II systems, we also visually rejected Na~I systems that had (a) the possibility of an interloping non-Na~I system, and (b) systems with low signal-to-noise within the doublet region.  Among the parent sample for Na~I systems, we claim 583 detections, 3796 questionable candidates, and 91272 non detections.  Many of the candidate systems were found at very low velocity and were affected by incomplete Na~I sky-line subtraction.  This turned out to be a severe problem.  To minimize contamination from incorrect sky subtraction, we excluded from the sample all candidate lines that fell between 5888.0~\AA~and 5901.0~\AA (S in Table~\ref{tab:parent}).  This reduced the number of Na~I detections down to 16 systems.  For the 16 detections, the line finder yields a mean line significance of ($\mu$,$\sigma$)=(2.8$\pm$0.2,0.7$\pm$0.3).  These 16 detections result from 15 unique galaxy spectra and 12 unique quasar spectra.   Listed in Table~\ref{tab:uniqNaI} are the individual Na~I absorber galaxies, with the same columns as in Table~\ref{tab:uniqCaII}.

\subsubsection{Ca~II~K and Na~I~D2 Sample}
We generated a third list in which both the Ca~II~K line and the Na~I~D2 line are detected by the line finder. In this list we use the Ca~II~K line as the ``confirmation" line for the Na~I~D2 line, and vice versa. This required that we impose a limit on the velocity agreement between the two species. We adopted a limit of 3 pixels, or 210~km~s$^{-1}$.  The visual inspection of this list, after applying the same sky subtraction processing as in the Na~I sample, yielded 2 detections.  The line finder gives a mean line significance for the Na~I lines of 2.8, and 2.3 for the Ca~II~K lines.  These 2 detections result from 2 unique galaxy and quasar spectra.   Listed in Table~\ref{tab:uniqboth} are the individual absorber galaxies.

In all Tables and Figures, `Na~I' refers to those absorbers in the Na~I Doublet Sample.   `Ca~II' refers to those absorbers in the Ca~II Doublet Sample.  `Both' refers to absorbers that have at least the Ca~II~K line and the Na~I~D2 line detected.  These absorbers are listed in the `Ca~II~K and Na~I~D2 Sample' and may also be listed in either the Ca~II or Na~I Doublet Sample. 
 
\subsection{Absorber Catalog}
The absorber catalog represents a merger of the above three final line lists of detections.  There may well be additional Ca~II and Na~I systems among the questionable candidates.  We investigated this possibility by stacking the spectra of questionable candidates using various cuts, and the doublets do become detectable in some of the stacks. Further analysis on the stacked spectra will be detailed in a future paper.  The confirmation of questionable candidates on an individual basis requires future spectroscopy with a higher signal-to-noise and spectral resolution.

As the line finder would occasionally mistake a Na~I blend for the Na~I~D2 line of the doublet, it would measure an equivalent width across the entire line profile, as opposed to just the Na~I~D2 component.  Due to this, we decided to remeasure the equivalent widths manually, to insure each line has a proper value.  To be consistent, we remeasured all the lines within the absorber sample.  We measured the rest equivalent width (rEW), equivalent width error, central wavelength, and central wavelength error, of the lines. The equivalent widths were computed from a direct summation of the pixels within the line region. The line equivalent width error is the total error of the sum of the pixel equivalent width errors.  The central wavelength of a line is the equivalent-width weighted mean of the pixels' wavelengths; its errors is the uncertainty of the mean.  

Table~\ref{tab:posinfo} is our absorber catalog\footnote[1]{This catalog does not include Mrk~1456 and SDSS J211701.26-002633.7 \citep{cherinka09}.  Mrk~1456 was selected from a visual inspection of the SDSS-DR3 spectrum of its background quasar. Here (DR7) it remained in the questionable candidate list of Ca~II doublets. SDSS~J211701.26-002633.7 was selected from the SDSS-DR5 using the same automated line finder but with slightly different parameters. Here it also remained in the questionable candidate list of Ca~II doublets.}. 
The columns of the catalog are: galaxy spectrum IAU designation, galaxy redshift, galaxy Galactic coordinates, quasar spectrum IAU designation, quasar redshift, quasar Galactic coordinates, impact parameter, the ratio of impact parameter to r-band Petrosian radius, and an identifier for which absorber list it belongs to. 

Table~\ref{tab:lineinfo} is a list of the measured line parameters for each absorber.  The columns are galaxy spectrum, SDSS redshift, the catalog the absorber belongs to, the Local Standard of Rest (LSR) velocity along the sightline to the quasar, then for Line 1 in the doublet: the rEW of the line, the significance of the line (rEW/$\sigma_{rEW}$), the heliocentric redshift of the line, and the velocity offset of the line from that of the expected position.  For Line 2, we list the rEW of the line, the heliocentric redshift of the line, and the velocity offset of the line from that of the expected position.  For blended Na~I features, Line 1 reports the blended values, while Line 2 is blank.   

In Figures~\ref{fig:caiiabs}-\ref{fig:bothabs}, we show a few examples of absorbers from the `Ca~II', `Na~I', and `Both' samples, ordered by IAU galaxy name.  All absorbers shown were selected using the LSR velocity cutoff (see Section~4.2).  Each figure displays an image centered on the galaxy, with the image size given in the lower left.  The galaxy name, redshift and impact parameter to the quasar are also displayed.  If located within the image, the quasar is marked with a white arrow.  All objects with spectroscopy are indicated with a red square.  To the right of each image displays the Ca~II or Na~I doublet lines, centered, in velocity space, on the redshift of the galaxy.  The dotted line indicates the velocity offset of the absorber from the galaxy position.

\section{Analysis of the Absorbers}
This absorber catalog consists of 108 pairs of galaxy-quasar spectra with either the Ca~II or Na~I doublet detected, or both.  The absorbers can be broadly classified into Galactic or extra-galactic systems.  The selection of these systems may vary depending on the science one wishes to explore through follow-up observations.   

\subsection{Selection from Equivalent Width Significance}
  Figure~\ref{fig:rewvssig} shows the significance of the equivalent width measurements as a function of the rest equivalent width of the strong line in a doublet (Ca~II~K or Na~I~D2).  The 3$\sigma$ limit, or 99$\%$ confidence limit, is indicated as a red horizontal line in the figure.   We find 43 systems with the significance of either Ca~II~K or Na~I~D2 line $>$ 3.0, 53 systems with a significance between 2-3, and 12 systems with a significance $<$2.0.  While we set the line finder to only accept lines with a significance $>$ 2, manually remeasuring the equivalent widths and errors has, in some instances, resulted in different significances than what the line finder reported.  This is due to having slightly different boundaries defining the line edges during the direct summation of the equivalent widths in each pixel.  
 
In the Milky Way, Ca~II and Na~I equivalent widths have measured values less than what has been observed in external galaxies.  Ca~II~K equivalent widths have been measured in the range of 0.08-0.8~\AA, with the median equivalent width around 0.18~\AA~\citep{bowen91b}.  \cite{welsh10} searched Milky Way sightlines (within 400~pc of the Sun) and found Ca~II~K  equivalent widths $<$0.3~\AA, and Na~I~D2 equivalent widths $<$0.5~\AA.  \cite{bowen91b} showed that the larger equivalent widths seen in external galaxies can be recovered from the smaller values seen in the Milky Way when taking into account contributions from multiple, unresolved components, as well as projection effects through a highly extended, inclined disc.  He showed that the equivalent widths seen in the Milky Way would be roughly doubled when viewed from outside our Galaxy.  

Equivalent widths, seen in nearby external galaxies, of Ca~II and Na~I range from 0.4-1.0~\AA~for the Ca~II~K line \citep{womble90,womble93,womble92,zych07,boks78,boks80,blades81,berg87,bowen91} and between $\sim$0.2-1.8~\AA~for the Na~I~D2 line \citep{junk94,womble90}, or $\sim$2.2~\AA~for blended Na~I systems \citep{kunth84}.   

There is overall agreement between our rEW values and those typically seen in the Milky Way and in nearby external galaxies.  As seen in Figure~\ref{fig:rewvssig}, we find that the Ca~II absorbers and the Na~I absorbers naturally divide, with the Ca~II absorbers located primarily at rEWs~$\lesssim$~0.5~\AA~and the Na~I absorbers primarily occupying the region with rEW~$\gtrsim$~0.5~\AA.  This division is possibly due to a bias between the Ca~II and Na~I samples, where the Ca~II systems include greater contributions from Galactic sightlines.  While both samples should include contributions from Galactic and extragalactic sightlines, the cut we placed on the Na~I sample to remove regions contaminated by incomplete Na~I sky subtraction, and hence, low velocity systems, may have effectively removed Galactic contributors.  Since this cut was not placed on the Ca~II sample, it becomes biased towards Galactic absorbers.  With this in mind, our Ca~II subset could be a mix of Galactic and extragalactic absorbers, weighted more towards Galactic contributions.  There are 49 Ca~II systems with rEW$>$0.3~\AA, that could be extragalactic in origin.  Our Na~I subset can also be a mix of Galactic and extragalactic systems, weighted more towards extragalactic systems.  All of our Na~I systems have a rEW$>$0.5~\AA, and thus could be considered extragalactic, based on their strength.  There are 2 Na~I systems that are resolved and have unusually high rEWs for what has been seen previously, but we have checked for interloping absorption systems and none could be identified.  

Figure~\ref{fig:rewvszgal} shows the rEW of our absorbers as a function of SDSS galaxy redshift.  We again see the split in rEW between the Ca~II and Na~I absorbers, as well as a separation in redshift.  Given the possible velocity overlap between Virgo Cluster galaxies and High-Velocity Clouds (see Section~4.2), if we consider the redshift region $\ge$0.01\footnote[2]{This redshift marks the estimate of Virgo's outer boundary \citep{bing93}.}, we find that the minimum rEW seen in the resulting absorbers is 0.5~\AA{}.  

While this cannot rule out extragalactic absorbers having rEW values below $\sim$0.5~\AA{}, this is consistent with rEW values seen in extragalactic absorbers so far.  The large cluster of Ca~II systems with rEW~$\lesssim$~0.5~\AA{} at low redshifts also points to a domination by contributions from Galactic sightlines.  

Examples of systems with rEW/$\sigma_{rEW}$ $>$ 3.0. are shown in Figures~\ref{fig:caiiabs}c, d for Ca~II, and Figures~\ref{fig:naiabs}e, f, g, k, n for Na~I.         

\subsection{Selection from LSR Velocity}
The absorber catalog contains a large number of low-redshift systems. Since we allowed velocity deviations of up to $\pm$500~km~s$^{-1}$ from the galaxies' redshifts, it is possible that some of our detections overlap with Ca~II and Na~I absorbers in the Milky Way Galaxy, with absorbers in the Local Group, or with absorbers in the Virgo Cluster.  One method of distinguishing Galactic absorbers from extragalactic ones is through an exploration of the v$_{LSR}$-galactic longitude-galactic latitude parameter space.  

\cite{wakker91} and \cite{bekhti08} have mapped out Galactic High-Velocity Cloud (HVC) Complexes around the Milky Way.  Figure~2b and 2e of \cite{wakker91} show v$_{LSR}$ vs {\it l} and {\it b}, respectively, of HVC's around the Milky Way.  The HVCs are not uniform across the sky.  They predominantly lie within two distinct regions.  One region of HVCs, with  -500$<$v$_{LSR}$$<$-90~km~s$^{-1}$, is located below the galactic plane ({\it b}~$<$~0$^{\circ}$) and at {\it l}~$<$~180$^{\circ}$.   Another group of HVCs predominantly lies in the region {\it l}~$<$~180$^{\circ}$, {\it b}~$>$~0$^{\circ}$, with LSR velocities between 90-345~km~s$^{-1}$.  The region of space between $v_{LSR}$$\pm$90~km~s$^{-1}$ is primarily located within the Galactic disk.  As most of the HVCs lie at $v_{LSR}$$<$345~km~s$^{-1}$, a cut above this limit will select out systems free from HVC contamination.  Below this cutoff, one must place tighter constraints on galactic {\it l} and {\it b}, or obtain higher-resolution data to disentangle likely contributions by Galactic gas to any absorber candidates.  

In Figure~\ref{fig:lsrvsl}, we show the LSR velocity of the Ca~II~K or Na~I~D2 absorbers as a function of Galactic latitude along the sightline towards the quasar.  Also plotted are lines indicating the $|v_{LSR}|$ $>$ 345~km~s$^{-1}$ cutoffs.  Most of our absorbers lie within the range that is easily confused with Galactic HVCs.  We find that the majority of Ca~II systems (also the ones located at small rEWs in Figure~\ref{fig:rewvssig}) lie in the space $v_{LSR}$$\pm$100~km~s$^{-1}$, and thus are most likely sightlines through the Galactic disk.  There are 4 Na~I systems around $v_{LSR}$$\sim$-200~km~s$^{-1}$ spread over 180$^{\circ}$ in Galactic longitude but are primarily concentrated around ${\it b}$$\sim$55$^{\circ}$.  These systems lie in an ${\it {l,b}}$ space empty of HVCs.  There are also 3 Ca~II systems located above the velocity cutoff for HVCs at $\sim$315$^{\circ}$.  Although these 7 systems lie in regions void of HVCs, there are a few Virgo cluster galaxies in our sample whose sightlines, due to our $\pm$500~km~s$^{-1}$ window on absorber line position, can scatter out to these LSR velocities.  These sightlines may then be either associated with the Virgo Cluster, the Galactic disk, or HVCs.  Above 600~km~s$^{-1}$, there are 11 Na~I systems and 3 Ca~II systems.  In general, we find the stronger Ca~II and Na~I equivalent width systems located at large LSR velocities, as expected if these systems were truly extragalactic in nature.  All the absorbers displayed in Figures~\ref{fig:caiiabs}-\ref{fig:bothabs} depict examples of systems with $|v_{LSR}|$$\ge$345~km~s$^{-1}$.     

\subsection{Selection from Impact Parameter}
An alternative to selecting on LSR velocity, one may also wish to select absorbers based on their impact parameter.  Figure~\ref{fig:bkpcvsrpet} shows the distribution of absorbers in impact parameter vs b/$r_{Petro}$ ratio.  The striping seen at low Petrosian radii is due to the large search radius (in angular scale) for extremely low redshift galaxies.  Each stripe corresponds to multiple quasar sightlines being associated with one galaxy.  As with the prior two selection criteria, we also find a natural division between the Ca~II and Na~I systems.  Most of the Ca~II systems seen at low Petrosian radii are due to the large search area around low-redshift galaxies picking up Galactic sightlines.  

DLAs are an absorber class of immense interest due to their large column densities and the fact that they trace the bulk of the neutral Hydrogen content in the Universe.  \cite{rao03} studied the properties of low-redshift DLA galaxies (0.05$<$z$<$1.0) and found them out to an impact parameter of 38~kpc, with a mean value of 12~kpc (converted to our cosmology).  \cite{rao11} recently identified a set of host galaxies within in the redshift range 0.1-1.0 exhibiting DLA, sub-DLA, and Lyman Limit (LL) absorption.  They found a median impact parameter of 17.4~kpc for the DLAs increasing to 36.4~kpc for the LL systems, with the full range extending out to 100~kpc.  Considering a mean impact parameter of 12~kpc, there are 3 systems found at small impact parameters that would be of interest.  For an impact parameter of 38~kpc, covering the entire range of DLA detections, there are 17 sightlines [9 galaxies] that are within the range where DLAs have been found.  

From HI maps of the Milky Way, as well as many nearby galaxies, we know the neutral Hydrogen extent of a galaxy can extend many times past the optical radius of the galaxy.  \cite{cayatte94,broeils97}  have shown that neutral Hydrogen can extend to at least twice the optical radius for spirals, and as much as 4-5x in dwarf galaxies \citep{swaters02,deblok96}.  It may be useful then to select systems based on the ratio of the impact parameter to the optical radius of the galaxy.  Here we use the r-band Petrosian radius as a proxy for the radius of the galaxy.  In some cases the Petrosian radius may miss extended light from the galaxy \citep{sdssedr}, so this radius can be thought of as a lower limit to the optical radius of the galaxy.  With this sample, we find 7 galaxies that lie within 5x the Petrosian radius, 3 of which lie within the optical disk of the galaxy (b/$r_{Petro}$$<$1)).  

For extragalactic systems, the median impact parameter seen for Ca~II and Na~I is $\sim$15~kpc, and $\sim$14~kpc, respectively \citep{womble90,womble93,womble92,zych07,boks78,boks80,blades81,berg87,bowen91,junk94}.  Considering this area of parameter space in Figure~\ref{fig:bkpcvsrpet}, we find our Ca~II and Na~I absorbers are located close in to galaxies with Petrosian radii $\sim$10~kpc, the median value seen in the SDSS~Spec-Photo sample we draw from.  These systems then are most likely extragalactic.  Examples of systems with b/$r_{Petro}$$<$5 are shown in Figures~\ref{fig:caiiabs}a, c, d, for Ca~II and Figures~\ref{fig:naiabs}a, e, f, g, h for Na~I.  

\subsection{Notes on Individual Absorbers}
Here we provide a few notes on individual absorbers from our absorber list.  

\begin{itemize}
 {\bf \item SDSS~J102703.86+283721.9 \\ }
This galaxy spectrum has a SDSS redshift of 0.00020$\pm$0.00016, classified manually with high confidence, and has a zwarning of `Absorption lines inconsistent'.  The galaxy is not associated with NGC~3245A, which has a redshift of 0.00441$\pm$0.000008.  Within the 100~kpc search radius, seven quasar sightlines were identified with Ca~II absorption, and one sightline with Na~I absorption.  However, the uncertainty of this galaxy's redshift may have resulted in erroneous matches and calls into question the apparent association of this galaxy with these sightlines.  We exclude this galaxy from Figures~\ref{fig:bkpcvsrpet} and \ref{fig:grMr}.  

{\bf \item SDSS~J113420.50-033525.4 \\ }
This galaxy has a SDSS redshift of 0.00008$\pm$0.00007, at high confidence.  Twelve quasars sightlines were identified with Ca~II absorption and four sightlines with Na~I absorption.  Upon viewing the spectrum for this galaxy, there appear to be several absorption lines indicating a stellar spectrum that corroborates the reported SDSS redshift.  Light from a star may have entered the fiber when the spectrum was taken.  This galaxy is also identified in the 2dFGRS as N171Z241, with a reported redshift of 0.1151$\pm$0.0003.  The spectrum for N171Z241 visually does not agree with the spectrum from SDSS.  Although the sightlines discovered through this association are real detections of Na~I or Ca~II, this discrepancy in the galaxy redshift leaves some doubt as to the associations between this galaxy with these quasar sightlines.  We exclude this galaxy from Figures~\ref{fig:bkpcvsrpet} and \ref{fig:grMr}.  

{\bf \item SDSS~J114313.05+193646.9 \\ }
The galaxy lies at a redshift of 0.02085$\pm$0.00009.  There is one quasar (SDSS~J114323.71+193448.0) located 81~kpc from the galaxy position.  This galaxy is in the cluster Abell~1367 (as indicated in NED\footnote[3]{NASA Extragalactic Database}), along with SDSS J114318.07+193401.3 and SDSS J114336.98+193616.7.  As these three galaxies intersect the same sightline to the quasar  SDSS~J114323.71+193448.0, it is unknown which galaxy is responsible for the absorbing gas, or if the gas is intracluster material.

{\bf \item SDSS~J114318.07+193401.3 \\ }
The galaxy lies at a redshift of 0.02262$\pm$0.00015, and is located 42~kpc from the quasar.  This galaxy is in the cluster Abell~1367.    

{\bf \item SDSS~J114336.98+193616.7 \\ }
The galaxy lies at a redshift of 0.02201$\pm$0.00019, and is located 92~kpc from the quasar.  This galaxy is in the cluster Abell~1367.    

{\bf \item SDSS~J150400.94+240437.1 \\ }
The galaxy lies at a redshift of 0.06894$\pm$0.00010.  The quasar, SDSS~J150359.48+240532.8, is located 78~kpc from the galaxy position.  This galaxy is possibly in a group with SDSS~J150403.17+240559.8.  The two galaxies intersect the same sightline to the quasar SDSS~J150359.48+240532.8, so it is unclear which galaxy is responsible for the absorber, or whether the absorption is coming from intragroup material.  

{\bf \item SDSS~J150403.17+240559.8 \\ }
The galaxy lies at a redshift of 0.06997$\pm$0.00019, and is located 77~kpc from the quasar.  This galaxy is in a group with SDSS~J150400.94+240437.1. 

\end{itemize}

\section{Characteristics of Extragalactic Absorber Galaxies}
In the previous section we described various ways to differentiate Galactic absorbers from extragalactic ones.  While the absorbers in the Galactic disk or in HVCs are interesting in and of themselves, our original purpose was to identify Ca~II and Na~I systems due to, as well as residing close to, nearby external galaxies. 

\subsection{Emission-Selected Systems}
Considering the criteria b/$r_{Petro}$~$<$~5, $|v_{LSR}|$~$\ge$~345~km~s$^{-1}$, and rEW/$\sigma_{rEW}$~$\ge$~3, there are 4 absorption systems that satisfy all three conditions.  There are two systems detected in Ca~II, one of which was also detected in Na~I, and two detected in Na~I.  These systems are SDSS~J141745.62+162509.3 (Figure~\ref{fig:caiiabs}c), SDSS~J155752.51+342142.8 (Figures~\ref{fig:caiiabs}d and \ref{fig:bothabs}b), SDSS~J122037.63+283803.3 (Figure~\ref{fig:naiabs}f), and SDSS~J140613.22+153035.5 (Figure~\ref{fig:naiabs}g).  These 4 systems have a mean g-r color of 0.6$\pm$0.2 and a mean $M_r$ of -20.1$\pm$1.8.  Adopting an $^{0.1}r$-band $L^*$ of -21.28 \citep{blant03a}, we find an average $L^*_{r^{0.1}}$ of 0.8$\pm$0.4~L$_{\odot}$.  All colors and magnitudes were calculated using Kcorrect v4.2 \cite{blant07} and band-shifted to z=0.1, indicated with the 0.1 superscript.  The mean impact parameter of the 4 systems is 13.6$\pm$9.8~kpc, while the mean b/$r_{Petro}$ ratio is 1.0$\pm$0.6.     

Figure~\ref{fig:grMr} shows the galaxy rest-frame g-r color versus the SDSS absolute r-band magnitude, band-shifted to 0.1.  As one galaxy may have multiple quasar sightlines associated with it, we only plot the 22 unique galaxies.  The two galaxies with uncertain redshifts have been excluded from the plot.  Adopting a red/blue color separation from \cite{yan06}, we find 2 red galaxies and 20 blue galaxies.  In both the red and blue samples we find a mix of galaxies with Ca~II and Na~I sightlines.  Our four extragalactic Ca~II and Na~I systems lie in the lower right portion of the figure, and would be classified as blue, $\sim$$L_r^*$ galaxies.   

\subsection{Comparison with Absorption-Selected Systems}
\cite{zych07} selected Ca~II systems from the SDSS using the more traditional approach of identifying the absorption line in the quasar spectrum first, with follow-up imaging to locate the associated host galaxy.  They identified 5 Ca~II absorption systems in this manner and found them primarily located in blue, $\sim$L$^*$, star-forming galaxies with the Ca~II located at small impact parameters (11.6$\pm$8.4~kpc)\footnote[4]{Mean value calculated using data from their Table~5}.  Of the 5 pairs reported by \cite{zych07}, one is found in our sample, the galaxy SDSS~J111849.76-002109.9 (Figures~\ref{fig:caiiabs}a and \ref{fig:naiabs}a).  The Ca~II equivalent width we measure for SDSS~J111849.76-002109.9 is consistent with that reported in \cite{zych07}.  We also detect Na~I from this galaxy as well.  This galaxy, along with one additional Na~I system, is included in our list of extragalactic absorbers if we relax the condition on absorption line significance.  Of the other 4 galaxies in Zych's sample, one galaxy was detected within the fiber of the quasar and the other 3 do not have spectroscopy in SDSS, and thus are not in our sample of DR7 galaxies.  

It is interesting to note that there is a general agreement in the overall galaxy properties between our emission-selected absorbers and their absorption-selected absorbers.  Both methods returned blue, $\sim$L$^*$ galaxies.

\section{Summary}
We have searched the SDSS-DR7 for low-redshift galaxy-quasar pairs, where the quasar is projected within 100~kpc of the galaxy, and found 97489 galaxy/quasar pairs from a sample of 105783 spectroscopic quasars and 798948 spectroscopic galaxies.  From this list of pairs, we searched for Ca~II and Na~I absorbers, using an automated process followed by visual inspection, and found 108 absorption line systems, 92 of which are Ca~II, 16 of which are Na~I, along with $\sim$5600 questionable candidate features.  Within our sample of 108 absorbers, we find many sightlines that coincide with sightlines through the Galactic disk or through High-Velocity Cloud complexes in our halo.  

Our goal was to identify Ca~II and Na~I systems that lie within or near known low-redshift galaxies.  To this end, we placed constraints on the absorber line significance (rEW/$\sigma_{rEW}$ $\ge$3), on the LSR velocity ($|v_{LSR}|$$\ge$345~km~s$^{-1}$), and on the ratio of impact parameter to r-band Petrosian radius (b/$r_{Petro}$~$<$~5).  We find 4 bonafide extragalactic galaxies with Ca~II and/or Na~I absorption detected along sightlines to nearby quasars.  One is detected in Ca~II only, two are detected in Na~I only, and one is detected in both Ca~II and Na~I.  These absorption systems arise in blue, $\sim$$L_r^*$ galaxies.  With the absorbers' impact parameters within the range seen for low-redshift DLAs, these systems would be good candidates for follow-up HI 21~cm or UV observations to determine neutral Hydrogen column densities.  

Future work, forthcoming in Paper 2, will consist of a statistical exploration of different galaxy properties for the Ca~II and Na~I absorbers and non-absorbing systems via stacking techniques.  Another area of future work will be to compare the emission line properties of the absorbing galaxies in our sample with those from the absorption-selected samples.

%%----------------------------------------------------------------------------------------------------
     
\acknowledgments 
\indent  Funding for the SDSS and SDSS-II has been provided by the Alfred
P. Sloan Foundation, the Participating Institutions, the National
Science Foundation, the U.S. Department of Energy, the National
Aeronautics and Space Administration, the Japanese Monbukagakusho, the
Max Planck Society, and the Higher Education Funding Council for
England.  The SDSS Web Site is http://www.sdss.org/.  The SDSS is managed by the
Astrophysical Research Consortium for the Participating
Institutions. The Participating Institutions are the American Museum
of Natural History, Astrophysical Institute Potsdam, University of
Basel, University of Cambridge, Case Western Reserve University,
University of Chicago, Drexel University, Fermilab, the Institute for
Advanced Study, the Japan Participation Group, Johns Hopkins
University, the Joint Institute for Nuclear Astrophysics, the Kavli
Institute for Particle Astrophysics and Cosmology, the Korean
Scientist Group, the Chinese Academy of Sciences (LAMOST), Los Alamos
National Laboratory, the Max-Planck-Institute for Astronomy (MPIA),
the Max-Planck-Institute for Astrophysics (MPA), New Mexico State
University, Ohio State University, University of Pittsburgh,
University of Portsmouth, Princeton University, the United States
Naval Observatory, and the University of Washington. \\ 
\indent This research has made use of NASA's Astrophysics Data System
and of the NASA/IPAC Extragalactic Database (NED) which is operated by
the Jet Propulsion Laboratory, California Institute of Technology,
under contract with the National Aeronautics and Space Administration.  \\
\indent We acknowledge the help of Matt Chornick, a University of Pittsburgh physics undergraduate researcher.\\
\indent We would also like to thank Jeff Newman and Sandhya Rao for their help with some of the spectra.\\
\indent We acknowledge the usage of the HyperLeda database.

%\bibliography{references}

\begin{thebibliography}{}
\bibitem[Abazajian et al.(2009)]{sdssdr7} Abazajian, K.~N., et al.\ 2009, \apjs, 182, 543 
\bibitem[Baugh(2006)]{baugh06} Baugh, C.~M.\ 2006, Reports on Progress in Physics, 69, 3101 
\bibitem[Ben Bekhti et al.(2008)]{bekhti08} Ben Bekhti, N., Richter, P., Westmeier, T., \& Murphy, M.~T.\ 2008, \aap, 487, 583 
\bibitem[Benson(2010)]{benson10} Benson, A.~J.\ 2010, \physrep, 495, 33 
\bibitem[Bergeron et al.(1987)]{berg87} Bergeron, J., Kunth, D., \& D'Odorico, S.\ 1987, \aap, 180, 1 
%\bibitem[Bertin et al.(1993)]{bertin93} Bertin, P., Lallement, R., Ferlet, R., \& Vidal-Madjar, A.\ 1993, \aap, 278, 549 
\bibitem[Binggeli et al.(1993)]{bing93} Binggeli, B., Popescu, C.~C., \& Tammann, G.~A.\ 1993, \aaps, 98, 275 
\bibitem[Blades et al.(1981)]{blades81} Blades, J.~C., Hunstead, R.~W., \& Murdoch, H.~S.\ 1981, \mnras, 194, 669 
\bibitem[Blanton et al.(2003)]{blant03a} Blanton, M.~R., et al.\ 2003, \apj, 592, 819 
%\bibitem[Blanton et al.(2003)]{blant03b} Blanton, M.~R., et al.\ 2003, \apj, 594, 186 
\bibitem[Blanton et al.(2003)]{blant03} Blanton, M.~R., Lin, H., Lupton, R.~H., Maley, F.~M., Young, N., Zehavi, I., \& Loveday, J.\ 2003, \aj, 125, 2276 
%\bibitem[Blanton et al.(2005)]{blanton05} Blanton, M.~R., Lupton, R.~H., Schlegel, D.~J., Strauss, M.~A., Brinkmann, J., Fukugita, M., \& Loveday, J.\ 2005, \apj, 631, 208 
\bibitem[Blanton \& Roweis(2007)]{blant07} Blanton, M.~R., \& Roweis, S.\ 2007, \aj, 133, 734
%\bibitem[Blanton \& Hogg(2009)]{blant09} Blanton, M.~R., \& Hogg, D., 2009, private communication
\bibitem[Boksenberg \& Sargent(1978)]{boks78} Boksenberg, A., \& Sargent, W.~L.~W.\ 1978, \apj, 220, 42 
\bibitem[Boksenberg et al.(1980)]{boks80} Boksenberg, A., Danziger, I.~J., Fosbury, R.~A.~E., \& Goss, W.~M.\ 1980, \apjl, 242, L145 
\bibitem[Bowen et al.(1991)]{bowen91} Bowen, D.~V., Pettini, M., Penston, M.~V., \& Blades, C.\ 1991, \mnras, 249, 145 
\bibitem[Bowen(1991)]{bowen91b} Bowen, D.~V.\ 1991, \mnras, 251, 649 
\bibitem[Bowen et al.(2005)]{bowen05} Bowen, D.~V., Jenkins, E.~B., Pettini, M., \& Tripp, T.~M.\ 2005, \apj, 635, 880 
\bibitem[Broeils \& Rhee(1997)]{broeils97} Broeils, A.~H., \& Rhee, M.-H.\ 1997, \aap, 324, 877 
\bibitem[Bukhmastova(2001)]{bukhma00} Bukhmastova, Y.~L.\ 2001, Astronomy Reports, 45, 581 
\bibitem[Burbidge et al.(1990)]{burb90} Burbidge, G., Hewitt, A., Narlikar, J.~V., \& Gupta, P.~D.\ 1990, \apjs, 74, 675 
%\bibitem[Carilli \& van Gorkom(1987)]{carilli87} Carilli, C.~L., \& van Gorkom, J.~H.\ 1987, \apj, 319, 683 
%\bibitem[Carilli et al.(1989)]{carilli89} Carilli, C.~L., van Gorkom, J.~H., \& Stocke, J.~T.\ 1989, \nat, 338, 134 
\bibitem[Cayatte et al.(1994)]{cayatte94} Cayatte, V., Kotanyi, C., Balkowski, C., \& van Gorkom, J.~H.\ 1994, \aj, 107, 1003
%\bibitem[Chen \& Lanzetta(2003)]{chen03} Chen, H.-W., \& Lanzetta, K.~M.\ 2003, \apj, 597, 706
%\bibitem[Chen et al.(2010)]{chen10} Chen, Y.-M., Tremonti, C.~A., Heckman, T.~M., Kauffmann, G., Weiner, B.~J., Brinchmann, J., \& Wang, J.\ 2010, \aj, 140, 445 
\bibitem[Cherinka et al.(2009)]{cherinka09} Cherinka, B., Schulte-Ladbeck, R.~E., \& Rosenberg, J.~L.\ 2009, \aj, 138, 1714 
%\bibitem[Cherinka(0000)]{cherinka11} Cherinka, B., Thesis, in preparation.
%\bibitem[Christensen et al.(2007)]{christ} Christensen, L., Wisotzki, L., Roth, M.~M., S{\'a}nchez, S.~F., Kelz, A., \& Jahnke, K.\ 2007, \aap, 468, 587 
\bibitem[Churchill (0000)]{churchill} Churchill, C., in preparation, {\it The Analysis of Quasar Absorption Line Spectra}, \ Cambridge University Press
\bibitem[Crampton et al.(1997)]{crampton97} Crampton, D., Gussie, G., Cowley, A.~P., \& Schmidtke, P.~C.\ 1997, \aj, 114, 2353 
\bibitem[de Blok et al.(1996)]{deblok96} de Blok, W.~J.~G., 
McGaugh, S.~S., \& van der Hulst, J.~M.\ 1996, \mnras, 283, 18 
\bibitem[Dekel \& Birnboim(2006)]{dekel06} Dekel, A., \& Birnboim, Y.\ 2006, \mnras, 368, 2 
%\bibitem[Ferlet et al.(1985)]{ferlet85a} Ferlet, R., Dennefeld, M., \& Maurice, E.\ 1985, \aap, 152, 151 
%\bibitem[Ferlet et al.(1985)]{ferlet85} Ferlet, R., Vidal-Madjar, A., \& Gry, C.\ 1985, \apj, 298, 838 
%\bibitem[Frei \& Gunn(1994)]{frei95} Frei, Z., \& Gunn, J.~E.\ 1994,  \aj, 108, 1476
%\bibitem[Gardner et al.(2001)]{gardner} Gardner, J.~P., Katz, N., Hernquist, L., \& Weinberg, D.~H.\ 2001, \apj, 559, 131 
%\bibitem[Hewett et al.(1985)]{hewett85} Hewett, P.~C., Irwin, M.~J., Bunclark, P., Bridgeland, M.~T., Kibblewhite, E.~J., He, X.~T., \& Smith, M.~G.\ 1985, \mnras, 213, 971 
%\bibitem[Hewett \& Wild(2007)]{hewett07} Hewett, P.~C., \& Wild, V.\ 2007, \mnras, 379, 738 
%\bibitem[Hinshaw et al.(2009)]{hinshaw09} Hinshaw, G., et al.\ 2009, \apjs, 180, 225 
%\bibitem[Holmberg et al.(2006)]{holm06} Holmberg, J., Flynn, C., \& Portinari, L.\ 2006, \mnras, 367, 449
%\bibitem[Johansson \& Efstathiou(2006)]{johan06} Johansson, P.~H., \& Efstathiou, G.\ 2006, \mnras, 371, 1519 
\bibitem[Junkkarinen \& Barlow(1994)]{junk94} Junkkarinen, V.~T., \& Barlow, T.~A.\ 1994, Bulletin of the American Astronomical Society, 26, 1330 
%\bibitem[Kacprzak et al.(2008)]{kac08} Kacprzak, G.~G., Churchill, C.~W., Steidel, C.~C., \& Murphy, M.~T.\ 2008, \aj, 135, 922 
%\bibitem[Kacprzak et al.(2011)]{kac11} Kacprzak, G.~G., Churchill, C.~W., Evans, J.~L., Murphy, M.~T., \& Steidel, C.~C.\ 2011, arXiv:1106.3068 
%\bibitem[Kennicutt(1992)]{kenn92} Kennicutt, R.~C., Jr.\ 1992, \apjs, 79, 255 
\bibitem[Kere{\v s} et al.(2005)]{keres05} Kere{\v s}, D., Katz, N., Weinberg, D.~H., \& Dav{\'e}, R.\ 2005, \mnras, 363, 2 
\bibitem[Kere{\v s} et al.(2009)]{keres09} Kere{\v s}, D., Katz, N., Fardal, M., Dav{\'e}, R., \& Weinberg, D.~H.\ 2009, \mnras, 395, 160 
\bibitem[Kimm et al.(2011)]{kimm11} Kimm, T., Slyz, A., Devriendt, J., \& Pichon, C.\ 2011, \mnras, 413, L51 
%\bibitem[Kondo et al.(2006)]{kondo06} Kondo, S., et al.\ 2006, \apj, 643, 667 
\bibitem[Koz{\l}owski \& Kochanek(2009)]{kozlow09} Koz{\l}owski, S., \& Kochanek, C.~S.\ 2009, \apj, 701, 508 
\bibitem[Kunth \& Bergeron(1984)]{kunth84} Kunth, D., \& Bergeron, J.\ 1984, \mnras, 210, 873 
%\bibitem[Le Brun et al.(1997)]{lebrun97} Le Brun, V., Bergeron, J., Boisse, P., \& Deharveng, J.~M.\ 1997, \aap, 321, 733 
\bibitem[Makarov et al.(2003)]{mak03} Makarov, D.~I., Karachentsev, I.~D., \& Burenkov, A.~N.\ 2003, \aap, 405, 951 
%\bibitem[Marinoni et al.(1999)]{mar99} Marinoni, C., Monaco, P., Giuricin, G., \& Costantini, B.\ 1999, \apj, 521, 50
%\bibitem[M{\'e}nard et al.(2005)]{menard05} M{\'e}nard, B., 
%Zibetti, S., Nestor, D., \& Turnshek, D.\ 2005, IAU Colloq.~199: Probing Galaxies through Quasar Absorption Lines,86 
%\bibitem[M{\'e}nard \& Chelouche(2009)]{menard09} M{\'e}nard, B., \& Chelouche, D.\ 2009, \mnras, 393, 808 
%\bibitem[Nagamine et al.(2007)]{naga} Nagamine, K., Wolfe, A.~M., Hernquist, L., \& Springel, V.\ 2007, \apj, 660, 945 
%\bibitem[Nestor et al.(2008)]{nestor08} Nestor, D.~B., Pettini, M., Hewett, P.~C., Rao, S., \& Wild, V.\ 2008, \mnras, 1176 
%\bibitem[Nestor et al.(2011)]{nestor11} Nestor, D.~B., Johnson, B.~D., Wild, V., M{\'e}nard, B., Turnshek, D.~A., Rao, S., \& Pettini, M.\ 2011, \mnras, 412, 1559 
%\bibitem[Okoshi \& Nagashima(2005)]{okoshi} Okoshi, K., \& Nagashima, M.\ 2005, \apj, 623, 99 
%\bibitem[Park \& Choi(2005)]{park05} Park, C., \& Choi, Y.-Y.\ 2005, \apjl, 635, L29 
\bibitem[Paturel et al.(2003)]{paturel03} Paturel, G., Petit, C., Prugniel, P., Theureau, G., Rousseau, J., Brouty, M., Dubois, P., \& Cambr{\'e}sy, L.\ 2003, \aap, 412, 45 
\bibitem[Rao et al.(2003)]{rao03} Rao, S.~M., Nestor, D.~B., Turnshek, D.~A., Lane, W.~M., Monier, E.~M., \& Bergeron, J.\ 2003, \apj, 595, 94 
\bibitem[Rao et al.(2011)]{rao11} Rao, S.~M., Belfort-Mihalyi, M., Turnshek, D.~A., Monier, E.~M., Nestor, D.~B., \& Quider, A.~M.\ 2011, arXiv:1103.4047 
\bibitem[Reimers \& Hagen(1998)]{reimers98} Reimers, D., \& Hagen, H.-J.\ 1998, \aap, 329, L25 
%\bibitem[Richter et al.(2011)]{richter11} Richter, P., Krause, F., Fechner, C., Charlton, J.~C., \& Murphy, M.~T.\ 2011, \aap, 528, A12 
\bibitem[Rosenberg et al.(2006)]{rosen06} Rosenberg, J.~L., Bowen, D.~V., Tripp, T.~M., \& Brinks, E.\ 2006, \aj, 132, 478 
%\bibitem[Routly \& Spitzer(1952)]{routly} Routly, P.~M., \& Spitzer, L.~J.\ 1952, \apj, 115, 227 
%\bibitem[Rubin et al.(1985)]{rubin85} Rubin, V.~C., Burstein, D., Ford, W.~K., Jr., \& Thonnard, N.\ 1985, \apj, 289, 81
%\bibitem[Sandage \& Bedke(1988)]{atlas} Sandage, A., \& Bedke, J.\ 1988, NASA Special Publication, 496,  
%\bibitem[Savage \& Sembach(1996)]{savage96} Savage, B.~D., \& Sembach, K.~R.\ 1996, \araa, 34, 279 
%\bibitem[Schlegel et al.(1998)]{schlegel98} Schlegel, D.~J., Finkbeiner, D.~P., \& Davis, M.\ 1998, \apj, 500, 525 
\bibitem[Schneider et al.(1993)]{schneider93} Schneider, D.~P., et al.\ 1993, \apjs, 87, 45 
%\bibitem[Schneider et al.(2005)]{schneider05} Schneider, D.~P., et al.\ 2005, \aj, 130, 367 
\bibitem[Schneider et al.(2010)]{schneider10} Schneider, D.~P., et al.\ 2010, \aj, 139, 2360 
\bibitem[Schulte-Ladbeck et al.(2004)]{schulte04} Schulte-Ladbeck, R.~E., Rao, S.~M., Drozdovsky, I.~O., Turnshek, D.~A., Nestor, D.~B., \& Pettini, M.\ 2004, \apj, 600, 613 
\bibitem[Schulte-Ladbeck et al.(2005)]{schulte05} Schulte-Ladbeck, R.~E., K{\"o}nig, B., Miller, C.~J., Hopkins, A.~M., Drozdovsky, I.~O., Turnshek, D.~A., \& Hopp, U.\ 2005, \apjl, 625, L79 
%\bibitem[Shapley et al.(2001)]{shapley01} Shapley, A.~E., Steidel, C.~C., Adelberger, K.~L., Dickinson, M., Giavalisco, M., \& Pettini, M.\ 2001, \apj, 562, 95 
%\bibitem[Shapley et al.(2005)]{shapley05} Shapley, A.~E., Steidel, C.~C., Erb, D.~K., Reddy, N.~A., Adelberger, K.~L., Pettini, M., Barmby, P., \& Huang, J.\ 2005, \apj, 626, 698 
%\bibitem[Siluk \& Silk(1974)]{siluk74} Siluk, R.~S., \& Silk, J.\ 1974, \apj, 192, 51 
%\bibitem[Smith et al.(2002)]{smith02} Smith, J.~A., et al.\ 2002, \aj, 123, 2121
%\bibitem[Smoker et al.(2006)]{smoker06} Smoker, J.~V., Lynn, B.~B., Christian, D.~J., \& Keenan, F.~P.\ 2006, \mnras, 370, 151 
%\bibitem[Steidel et al.(1994)]{steidel94} Steidel, C.~C., Dickinson, M., \& Persson, S.~E.\ 1994, \apjl, 437, L75 
\bibitem[Steidel et al.(2010)]{steidel10} Steidel, C.~C., Erb, D.~K., Shapley, A.~E., Pettini, M., Reddy, N., Bogosavljevi{\'c}, M., Rudie, G.~C., \& Rakic, O.\ 2010, \apj, 717, 289 
\bibitem[Stewart et al.(2011)]{stewart11} Stewart, K.~R., Kaufmann, T., Bullock, J.~S., Barton, E.~J., Maller, A.~H., Diemand, J., \& Wadsley, J.\ 2011, \apjl, 735, L1 
\bibitem[Stoughton et al.(2002)]{sdssedr} Stoughton, C., et al.\ 2002, \aj, 123, 485 
%\bibitem[Strateva et al.(2001)]{strat01} Strateva, I., et al.\ 2001, \aj, 122, 1861 
%\bibitem[Strauss et al.(2002)]{strauss02} Strauss, M.~A., et al.\ 2002, \aj, 124, 1810 
\bibitem[Swaters et al.(2002)]{swaters02} Swaters, R.~A., van Albada, T.~S., van der Hulst, J.~M., \& Sancisi, R.\ 2002, \aap, 390, 829 
%\bibitem[Vallerga et al.(1993)]{vall93} Vallerga, J.~V., Vedder, P.~W., Craig, N., \& Welsh, B.~Y.\ 1993, \apj, 411, 729 
%\bibitem[Vladilo et al.(1993)]{vlad93} Vladilo, G., Molaro, P., Monai, S., D'Odorico, S., Ferlet, R., Madjar, A.~V., \& Dennefeld, M.\ 1993, \aap, 274, 37 
\bibitem[Yan et al.(2006)]{yan06} Yan, R., Newman, J.~A., 
Faber, S.~M., Konidaris, N., Koo, D., \& Davis, M.\ 2006, \apj, 648, 281 
\bibitem[Wakker \& van Woerden(1991)]{wakker91} Wakker, B.~P., \& van Woerden, H.\ 1991, \aap, 250, 509 
%\bibitem[Welsh et al.(2009)]{welsh09} Welsh, B.~Y., Wheatley, J., \& Lallement, R.\ 2009, \pasp, 121, 606 
\bibitem[Welsh et al.(2010)]{welsh10} Welsh, B.~Y., Lallement, R., Vergely, J.-L., \& Raimond, S.\ 2010, \aap, 510, A54 
%\bibitem[Welty et al.(1996)]{welty96} Welty, D.~E., Morton, D.~C., \& Hobbs, L.~M.\ 1996, \apjs, 106, 533 
%\bibitem[Wild \& Hewett(2005)]{wild05} Wild, V., \& Hewett, P.~C.\ 2005, \mnras, 361, L30 
%\bibitem[Wild et al.(2006)]{wild06} Wild, V., Hewett, P.~C., \& Pettini, M.\ 2006, \mnras, 367, 211 
%\bibitem[Wild et al.(2007)]{wild07} Wild, V., Hewett, P.~C., \& Pettini, M.\ 2007, \mnras, 374, 292 
\bibitem[Williams et al.(2005)]{will05} Williams, P.~R., Shu, C.-G., \& M{\'e}nard, B. \ 2005, IAU Colloq.~199: Probing Galaxies through Quasar Absorption Lines
\bibitem[Wolfe et al.(2005)]{wolf05} Wolfe, A.~M., Gawiser, E., \& Prochaska, J.~X.\ 2005, \araa, 43, 861 
\bibitem[Womble et al.(1990)]{womble90} Womble, D.~S., Junkkarinen, V.~T., Cohen, R.~D., \& Burbidge, E.~M.\ 1990, \aj, 100, 1785 
\bibitem[Womble et al.(1991)]{womble91} Womble, D.~S., Junkkarinen, V.~T., \& Burbidge, E.~M.\ 1991, \baas, 23, 1423 
\bibitem[Womble et al.(1992)]{womble92} Womble, D.~S., Junkkarinen, V.~T., \& Burbidge, E.~M.\ 1992, \apj, 388, 55 
\bibitem[Womble(1993)]{womble93} Womble, D.~S.\ 1993, \pasp, 105, 1043 
%\bibitem[Young \& Currie(1995)]{young95} Young, C.~K., \& Currie, M.~J.\ 1995, \mnras, 273, 1141 
%\bibitem[Zibetti et al.(2007)]{zibetti07} Zibetti, S., M{\'e}nard, B., Nestor, D.~B., Quider, A.~M., Rao, S.~M., \& Turnshek, D.~A.\ 2007, \apj, 658, 161 
%\bibitem[Zwaan et al.(2005)]{zwaan05} Zwaan, M.~A., van der Hulst, J.~M., Briggs, F.~H., Verheijen, M.~A.~W., \& Ryan-Weber, E.~V.\ 2005, \mnras, 364, 1467 
\bibitem[Zych et al.(2007)]{zych07} Zych, B.~J., Murphy, M.~T., Pettini, M., Hewett, P.~C., Ryan-Weber, E.~V., \& Ellison, S.~L.\ 2007, \mnras, 379, 1409 
%\bibitem[Zych et al.(2009)]{zych09} Zych, B.~J., Murphy, M.~T., Hewett, P.~C., \& Prochaska, J.~X.\ 2009, \mnras, 392, 1429 

\end{thebibliography}
%\bibliographystyle{plain}

%-------------------------------------------------------------------------
%% Figure and Table Area
\onecolumn

\clearpage

\clearpage

\clearpage
\begin{figure}[htb]
\centering
\subfloat[][]{\includegraphics[scale=0.5]{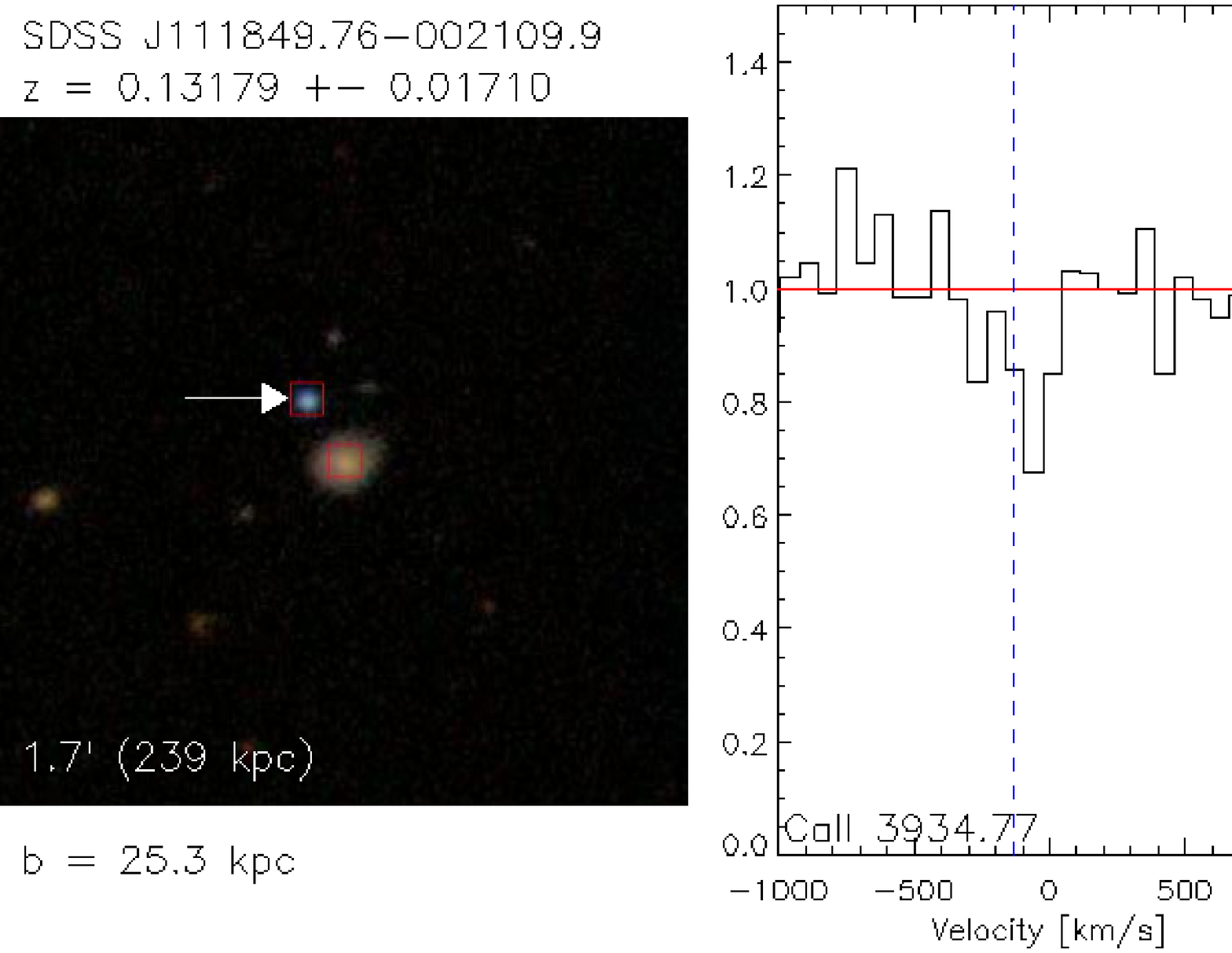}}
\qquad
\subfloat[][]{\includegraphics[scale=0.5]{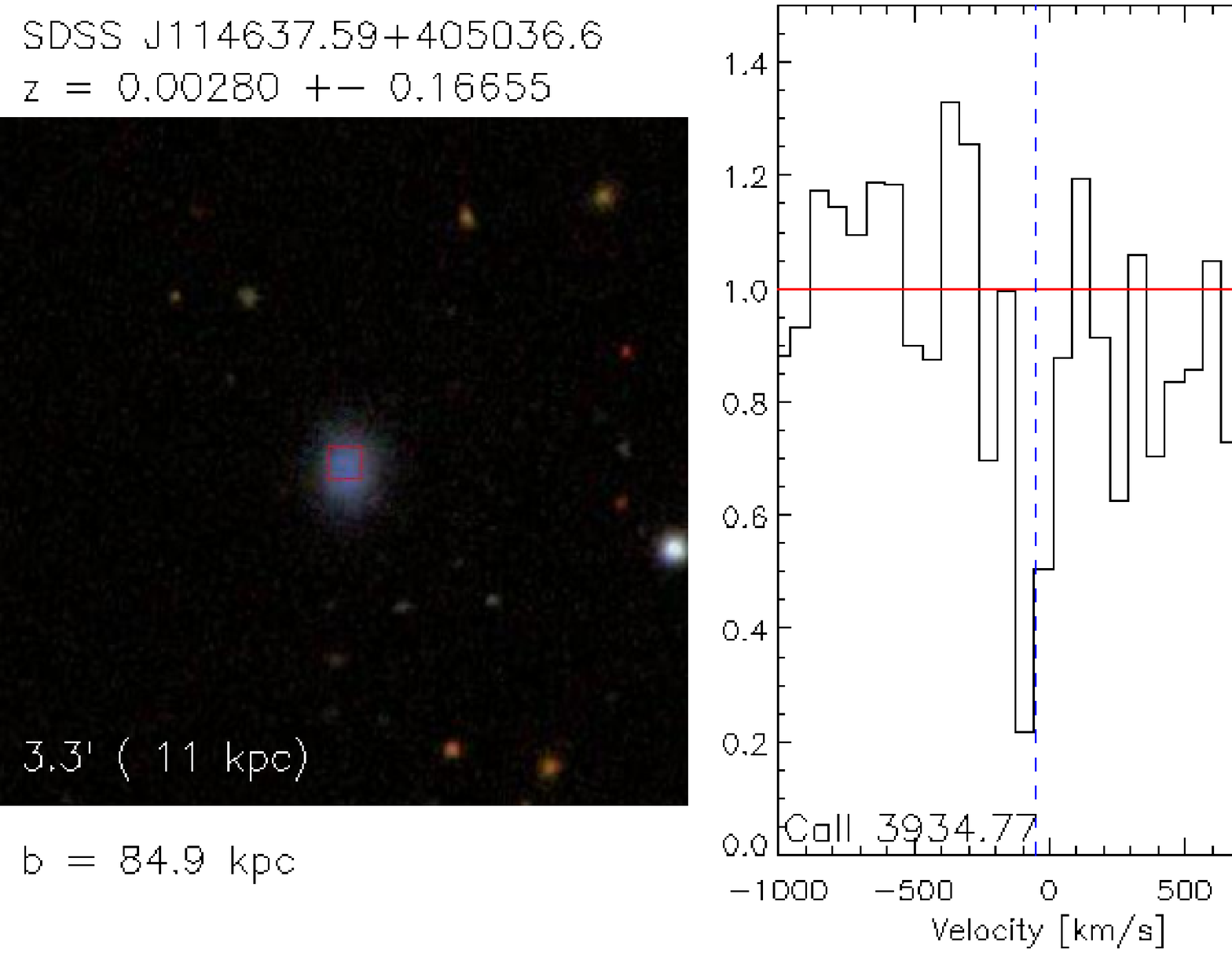}}
\caption{A subset of Ca~II Absorbers: The image is centered on the galaxy position, scaled to focus on the galaxy in each pair.  For cases where the QSO is within the image, it is marked with a white arrow.  In other cases, scaling the image to show both the galaxy and QSO results in an loss of detail regarding both objects.  All objects with an SDSS spectrum are marked with a red square.  The galaxy name and redshift are displayed at the top.  In the lower left is the image size.  Below that is the impact parameter between the quasar and galaxy.  The absorption lines detected are shown on the right.  The y-axis shows the normalized flux.  The middle panel displays the strong line at 3934.77\AA, with the weaker line at 3969.59\AA~on the right.  Each panel is centered in velocity space on the galaxy redshift.  The dashed blue line marks the velocity offset of the line from the host galaxy.  The red line marks the normalized continuum.}
\label{fig:caiiabs}
\end{figure}

\begin{figure}[htb]
\ContinuedFloat
\centering
\subfloat[][]{\includegraphics[scale=0.5]{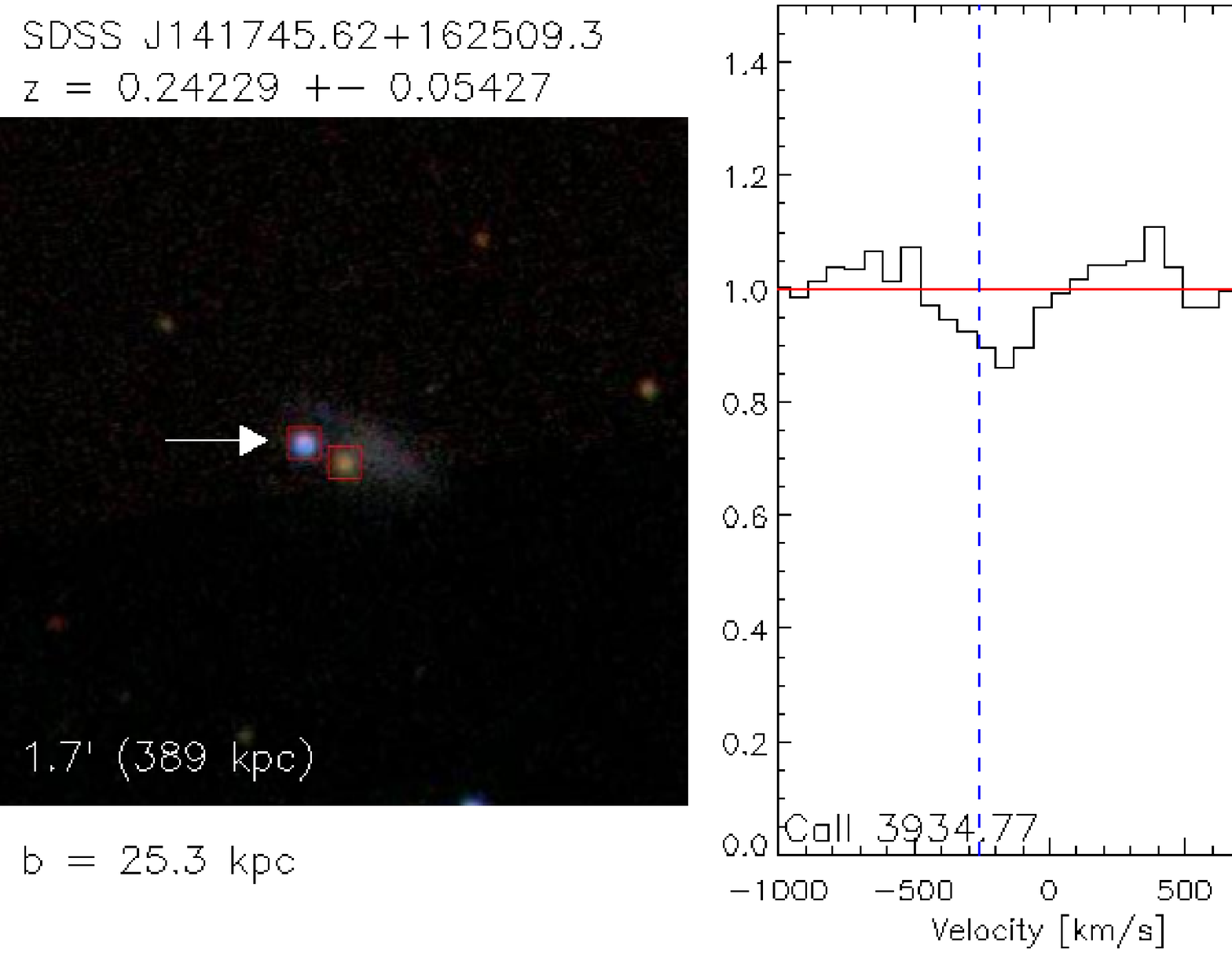}} 
\qquad
\subfloat[][]{\includegraphics[scale=0.5]{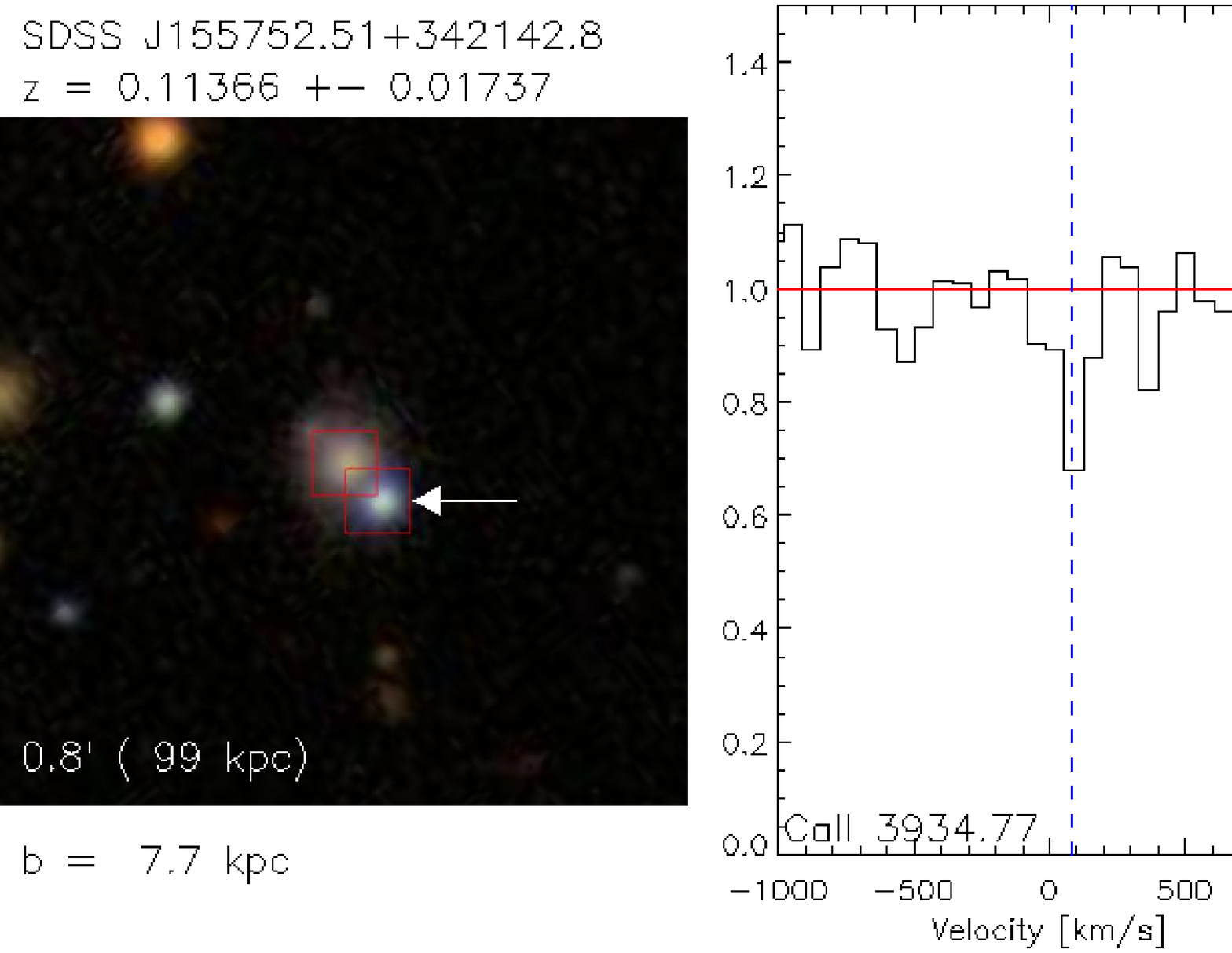}}
%\caption{Extragalactic Ca~II Absorbers:  same as in Figure~\ref{fig:caiiabs}}
\caption{cont'd}
%\label{fig:caiiabs2}
\end{figure}

\begin{figure}[htb]
\centering
\subfloat[][]{\includegraphics[scale=0.5]{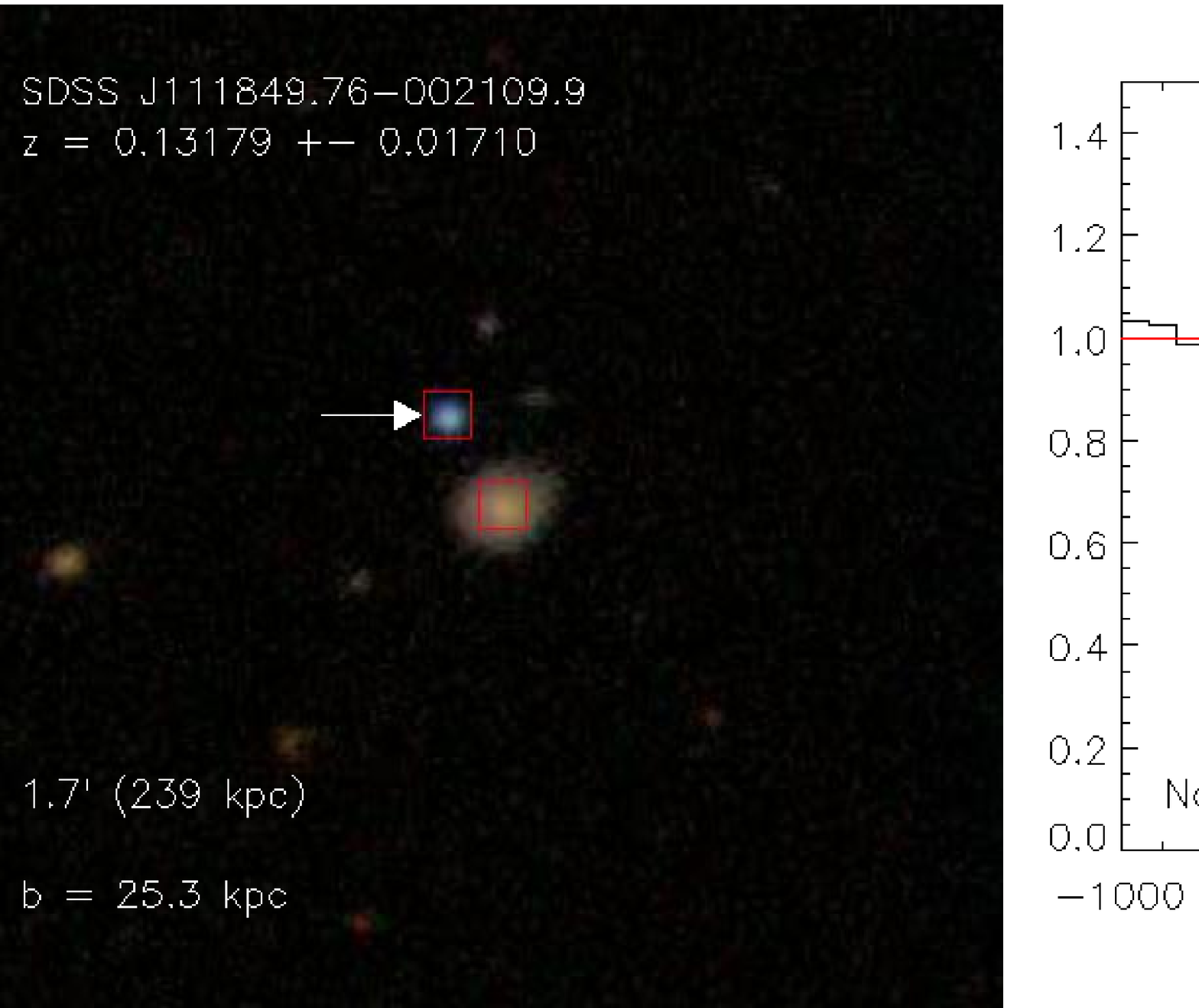}}
\qquad
\subfloat[][]{\includegraphics[scale=0.5]{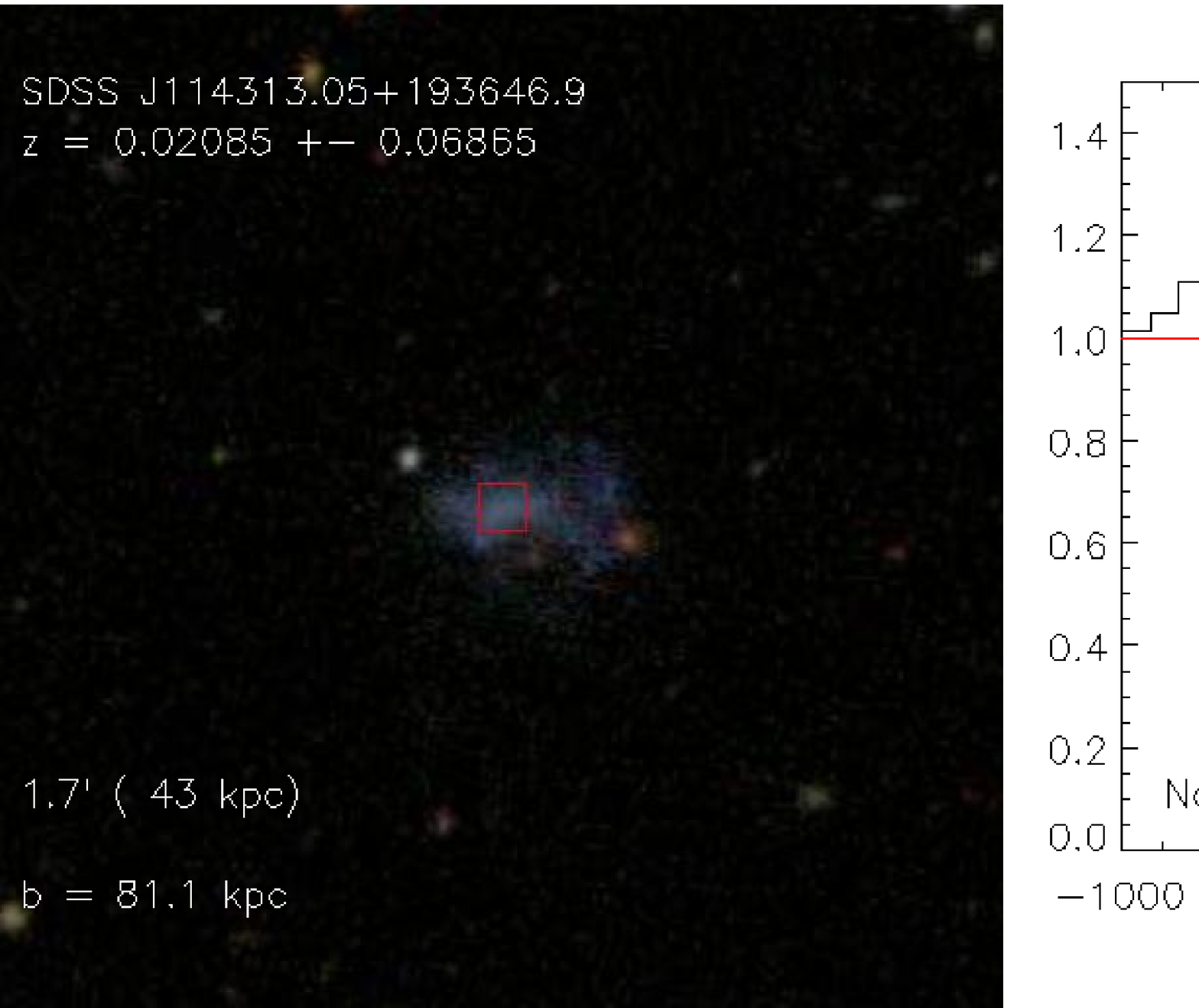}}
\caption{A subset of Na~I Absorbers:  The image marks are the same as in Figure~\ref{fig:caiiabs}.  The absorption line is shown in the right panel, centered on the expected position of the 5891.58~\AA~line from the galaxy redshift.  The y-axis shows the normalized flux.  The dashed blue line represents the velocity offset of the line from the expected position, or in the case of a blend, the offset as determined across the whole profile.}
\label{fig:naiabs}
\end{figure}

\begin{figure}[h]
\ContinuedFloat
\centering
\subfloat[][]{\includegraphics[scale=0.5]{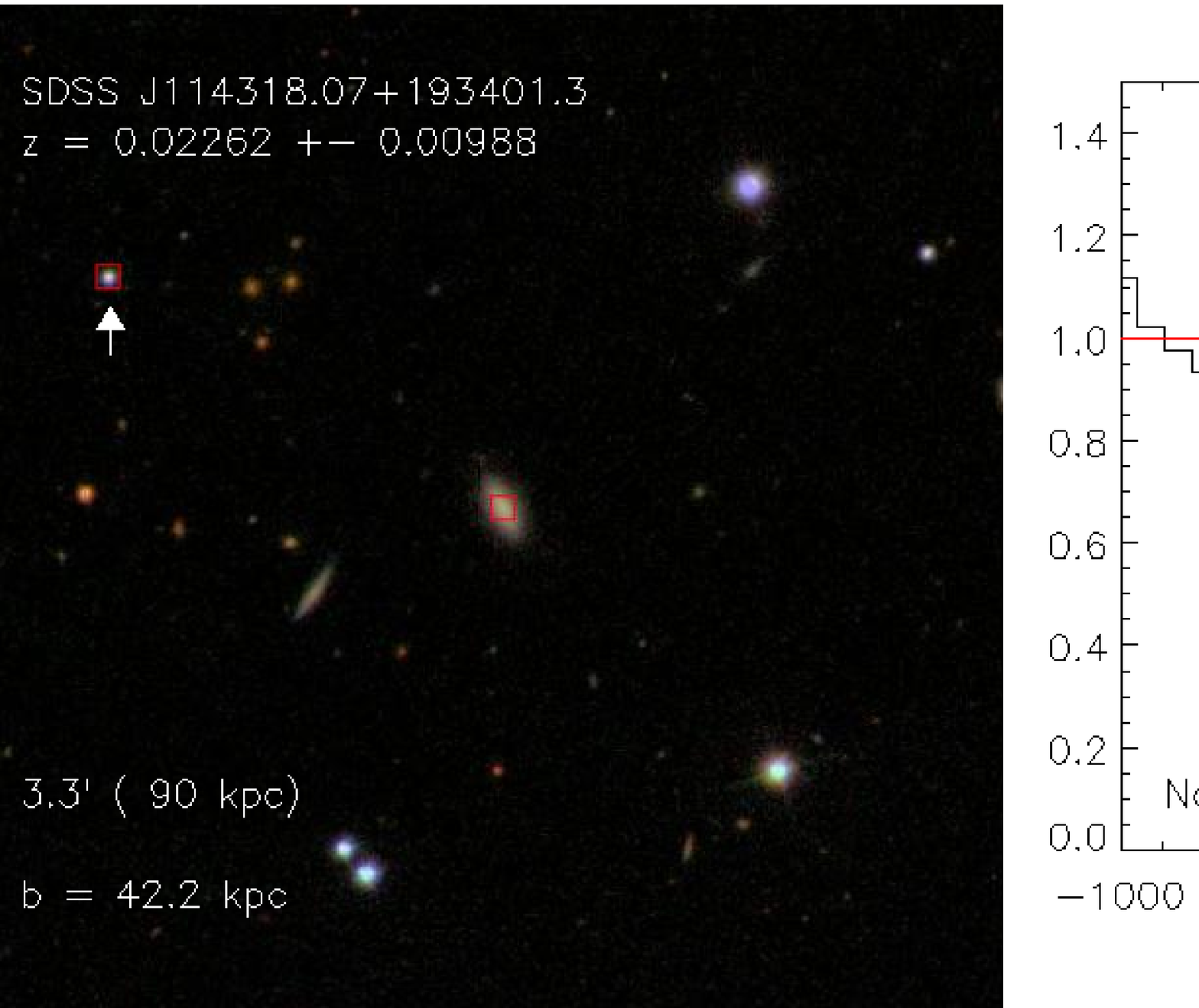}}
\qquad
\subfloat[][]{\includegraphics[scale=0.5]{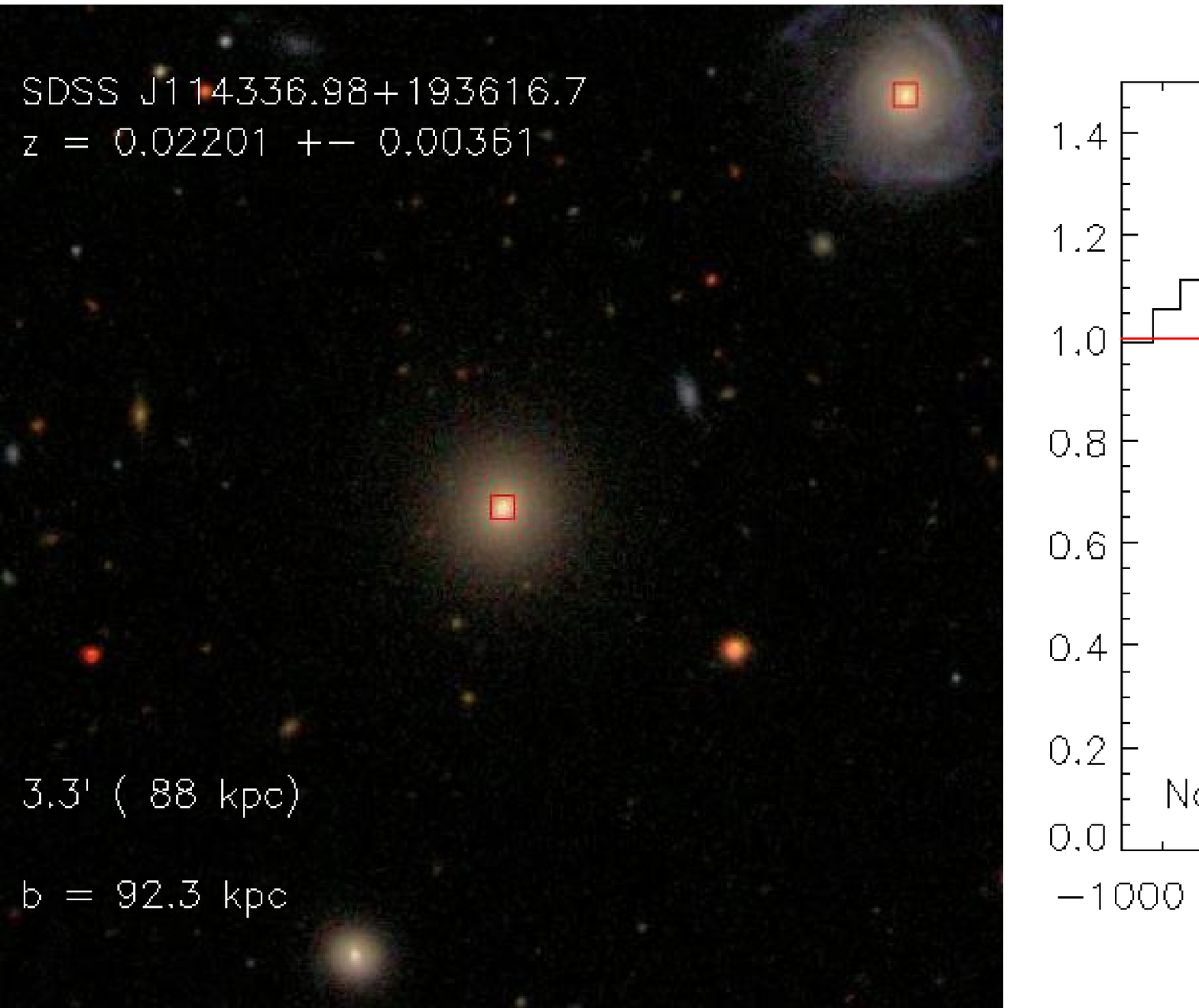}}
\caption{cont'd}
%\label{fig:naiabs2}
\end{figure}

\begin{figure}[h]
\ContinuedFloat
\centering
\subfloat[][]{\includegraphics[scale=0.5]{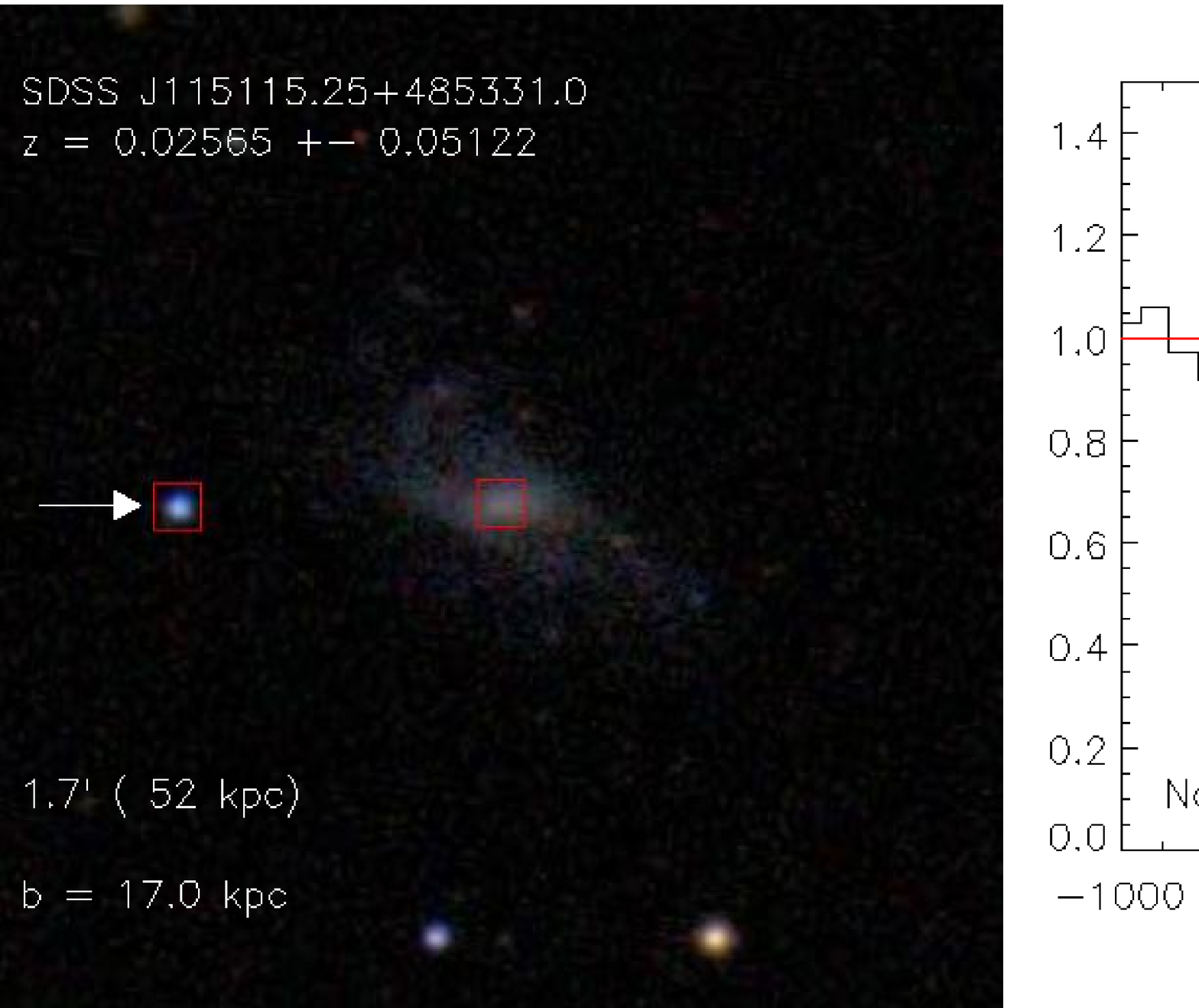}}
\qquad
\subfloat[][]{\includegraphics[scale=0.5]{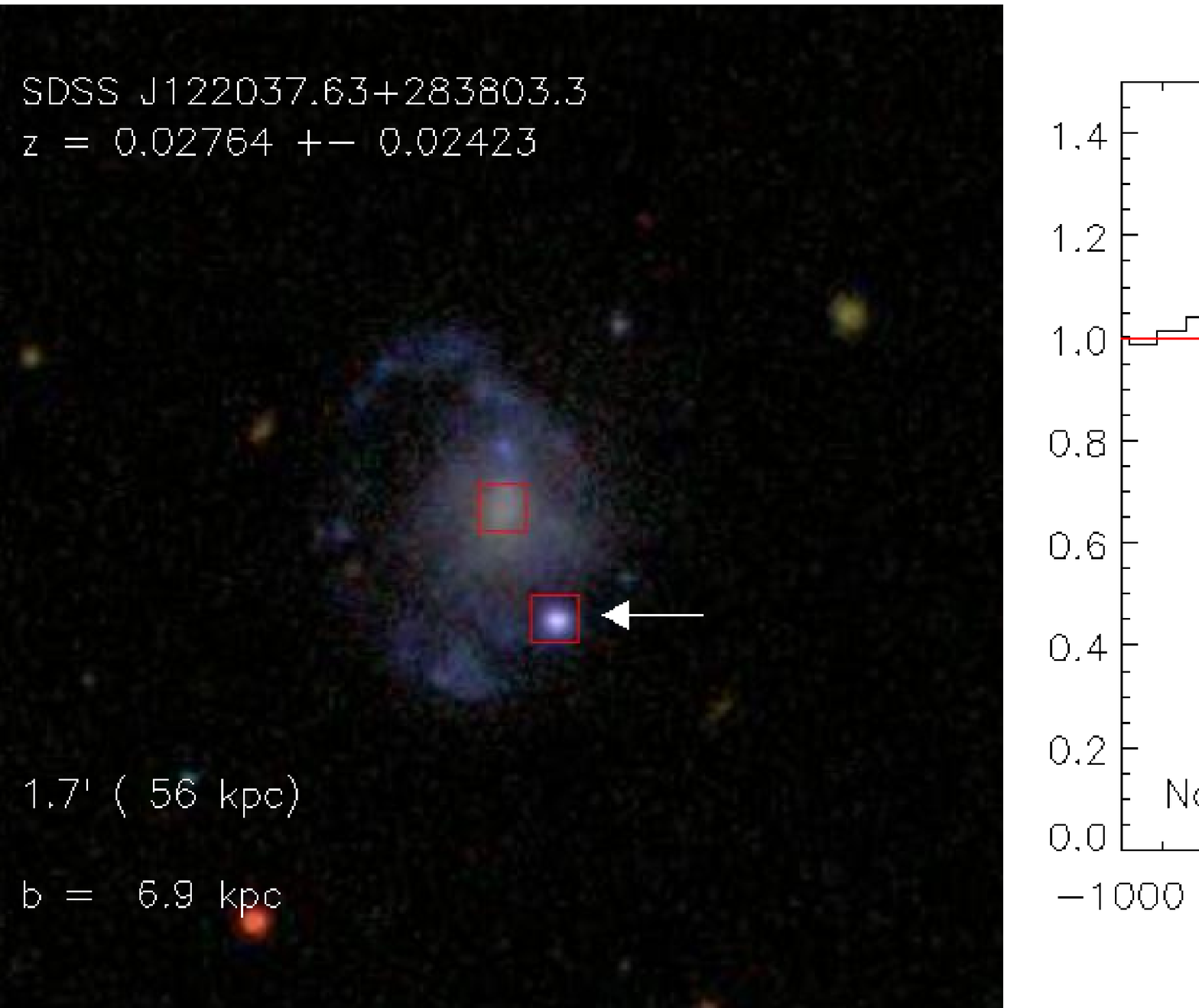}}
\caption{cont'd}
%\label{fig:naiabs3}
\end{figure}

\begin{figure}[h]
\ContinuedFloat
\centering
\subfloat[][]{\includegraphics[scale=0.5]{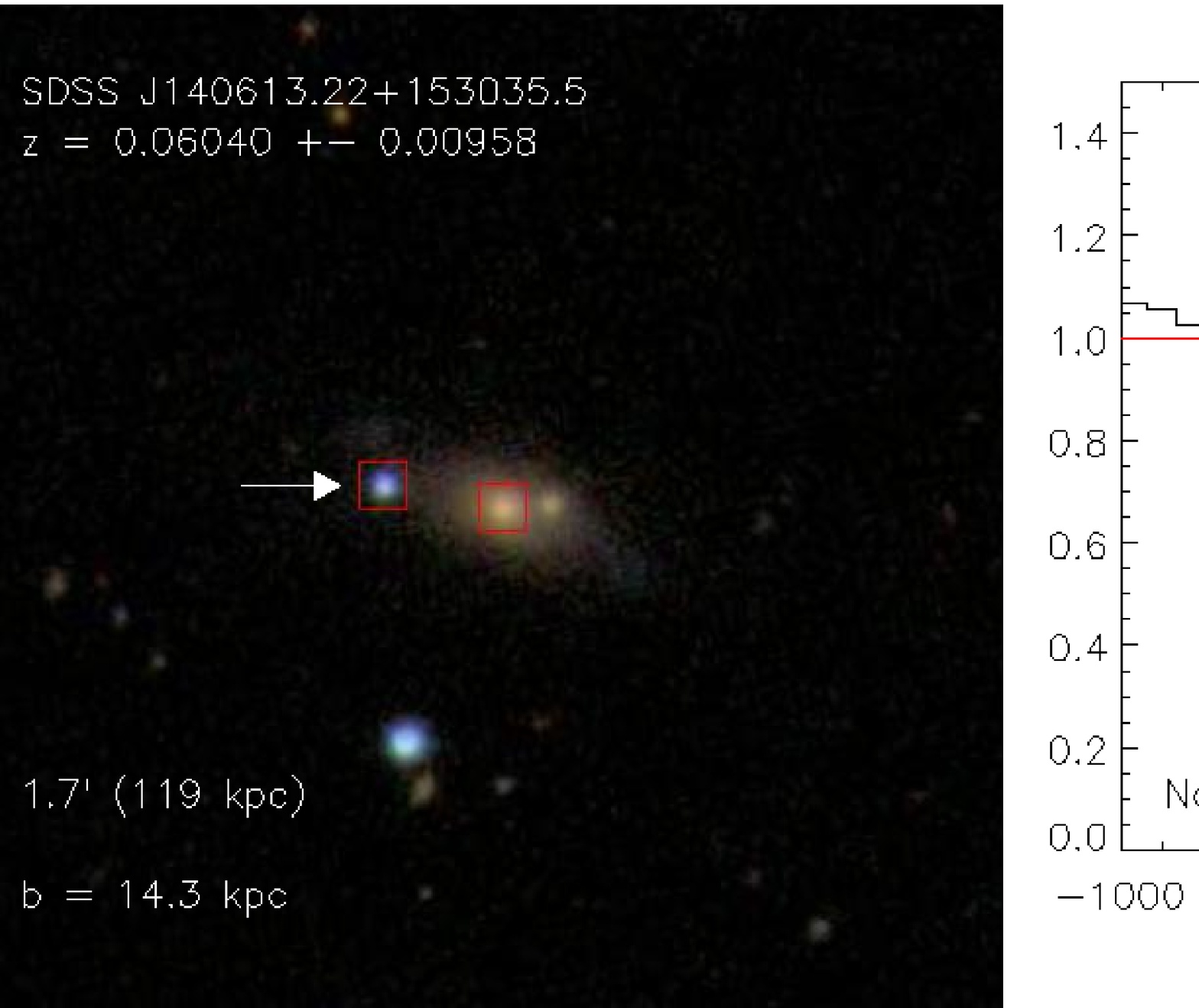}}
\qquad  
\subfloat[][]{\includegraphics[scale=0.5]{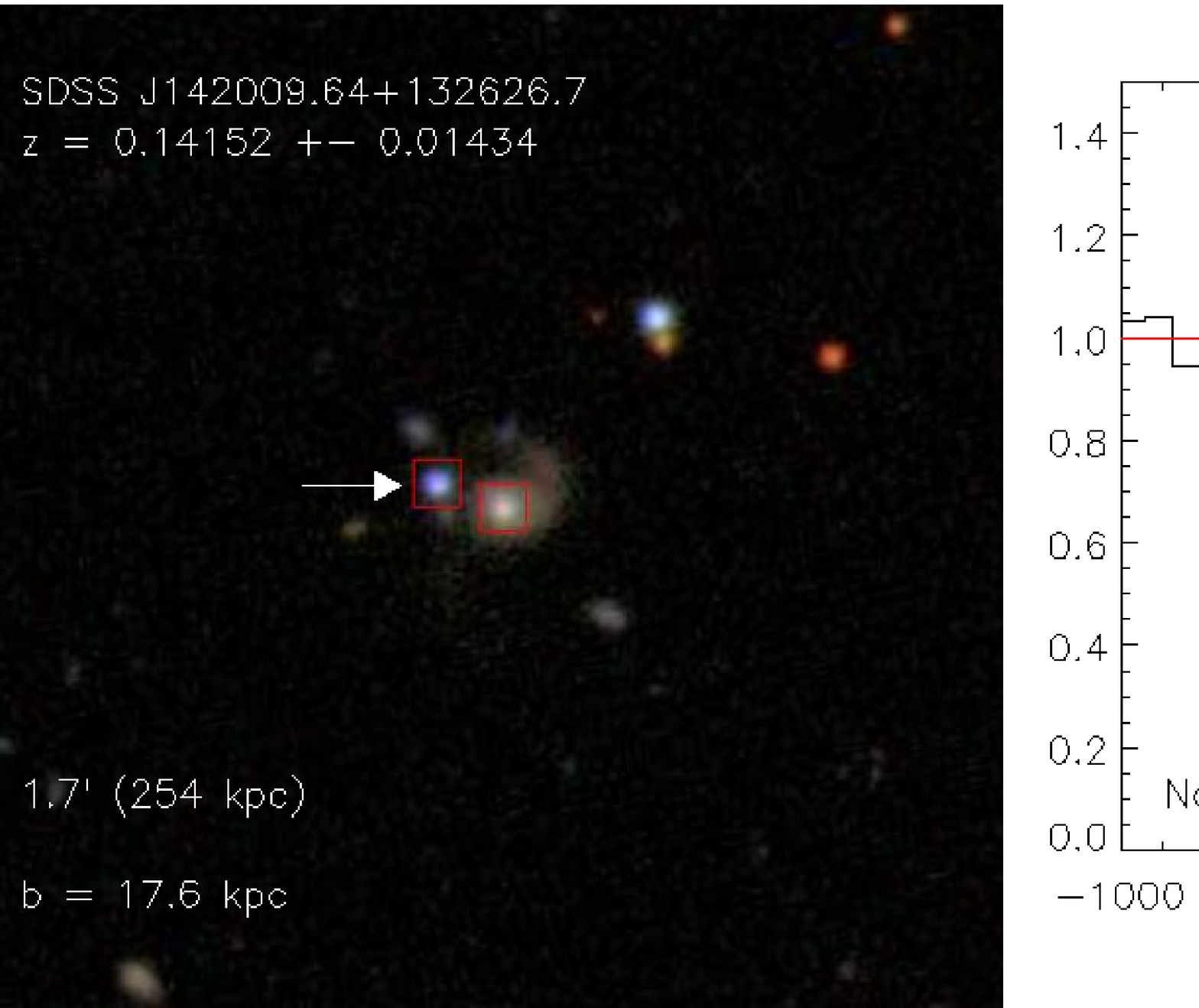}}
\caption{cont'd}
%\label{fig:naiabs4}
\end{figure}

\begin{figure}[h]
\ContinuedFloat
\centering
\subfloat[][]{\includegraphics[scale=0.5]{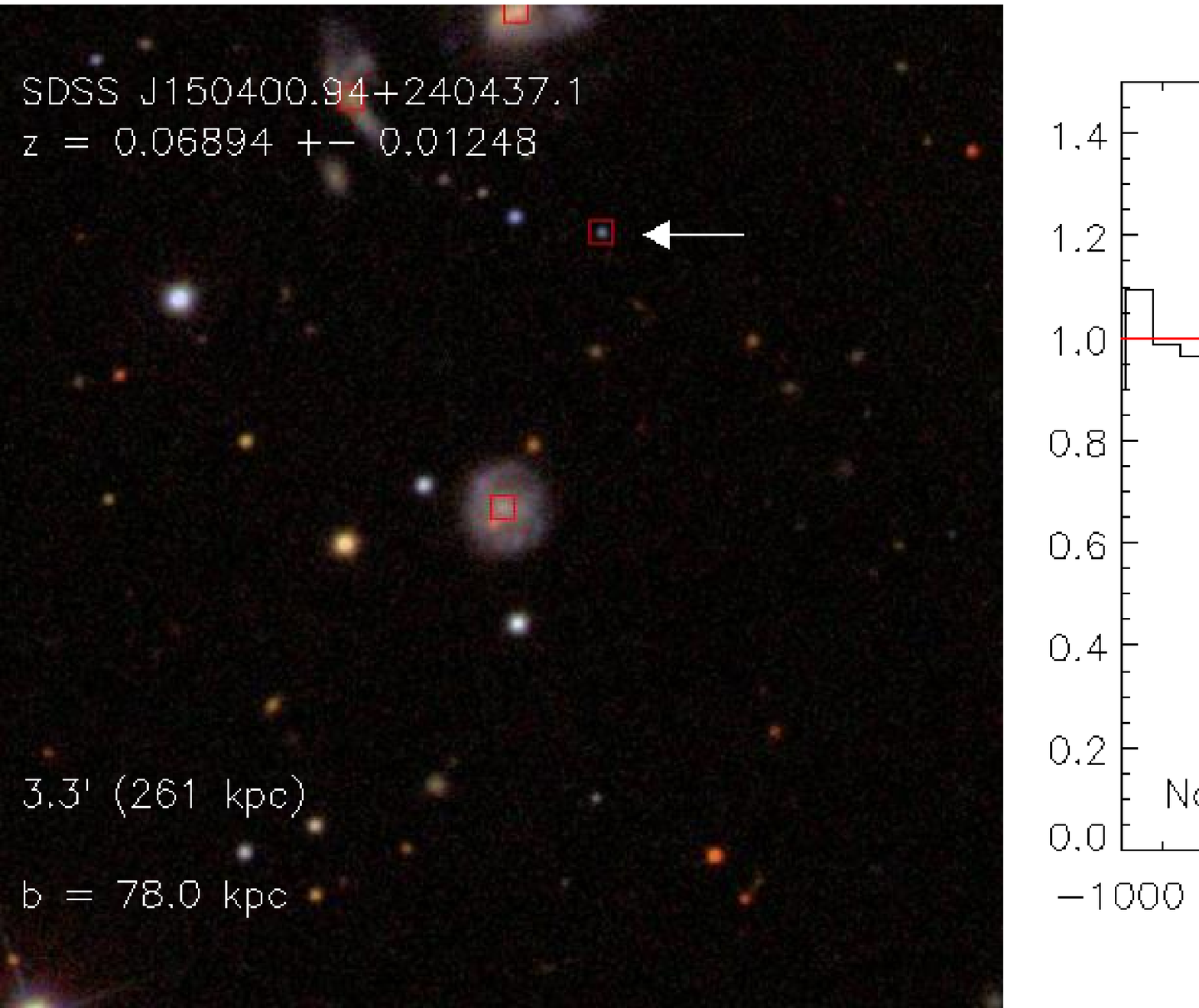}}
\qquad
\subfloat[][]{\includegraphics[scale=0.5]{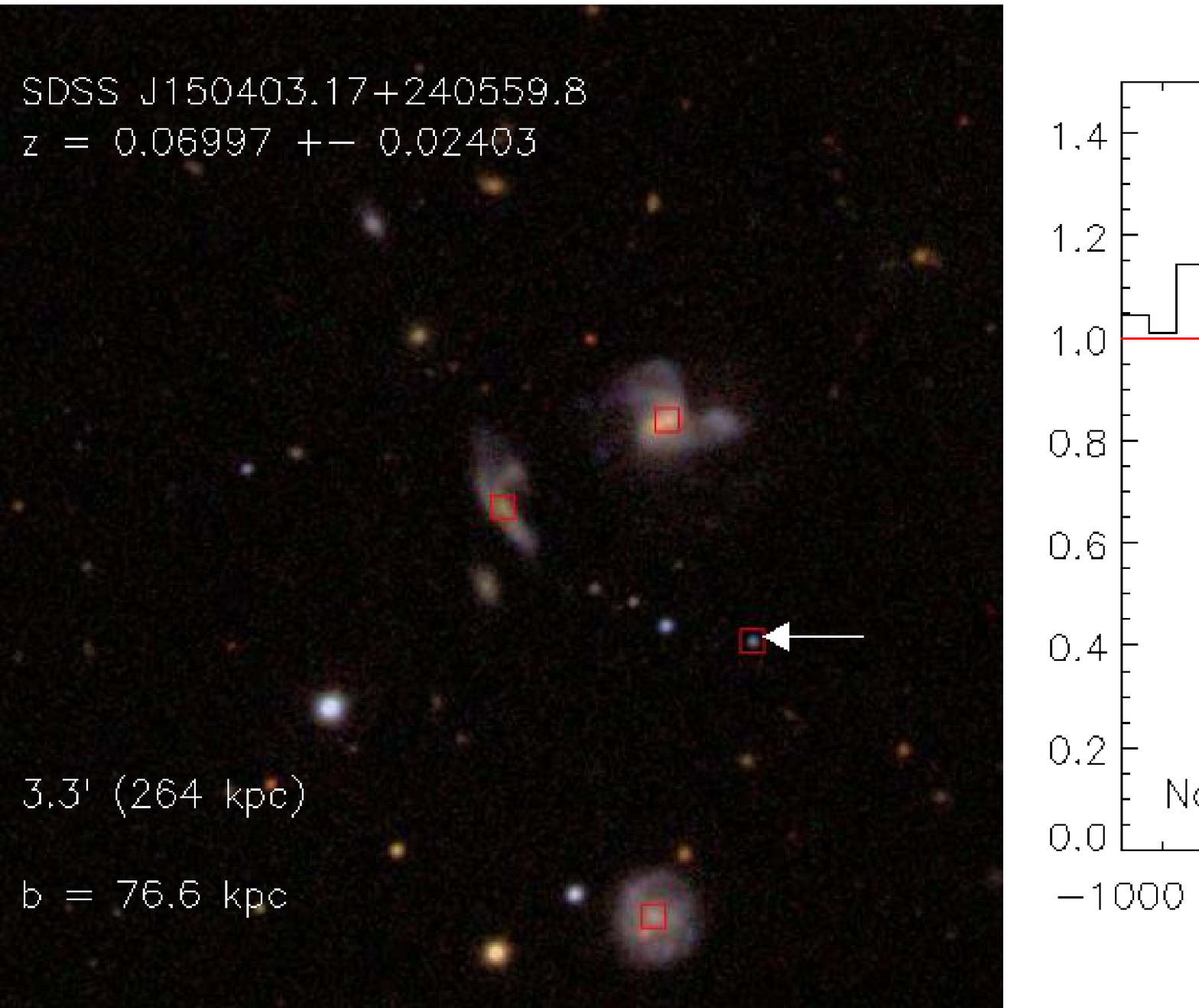}}
\caption{cont'd}
%\label{fig:naiabs5}
\end{figure}

\begin{figure}[h]
\ContinuedFloat
\centering
\subfloat[][]{\includegraphics[scale=0.5]{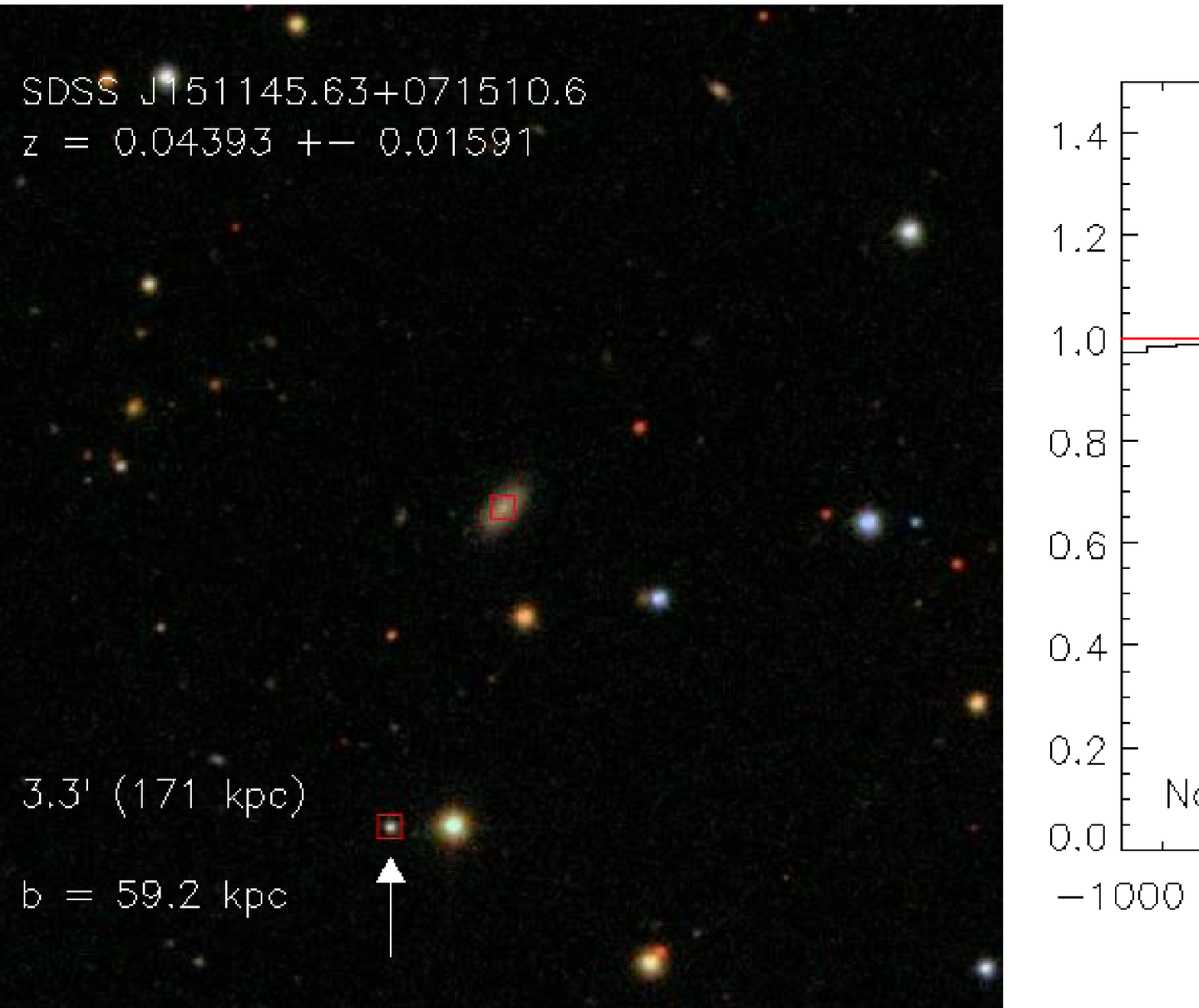}}
\caption{cont'd}
%\label{fig:naiabs6}
\end{figure}

\begin{figure}[h]
\centering
\subfloat[][]{\includegraphics[scale=0.5]{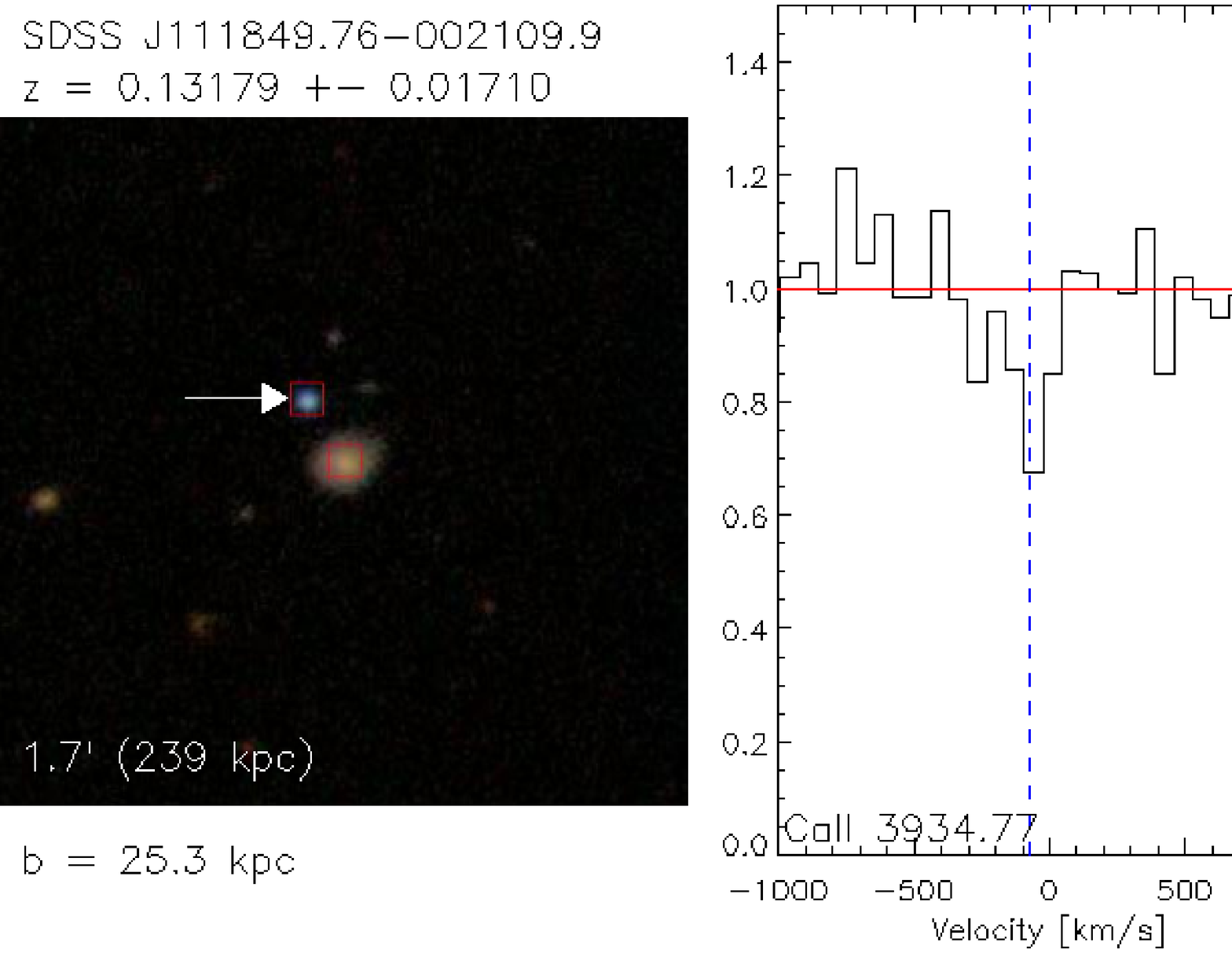}}
\qquad
\subfloat[][]{\includegraphics[scale=0.5]{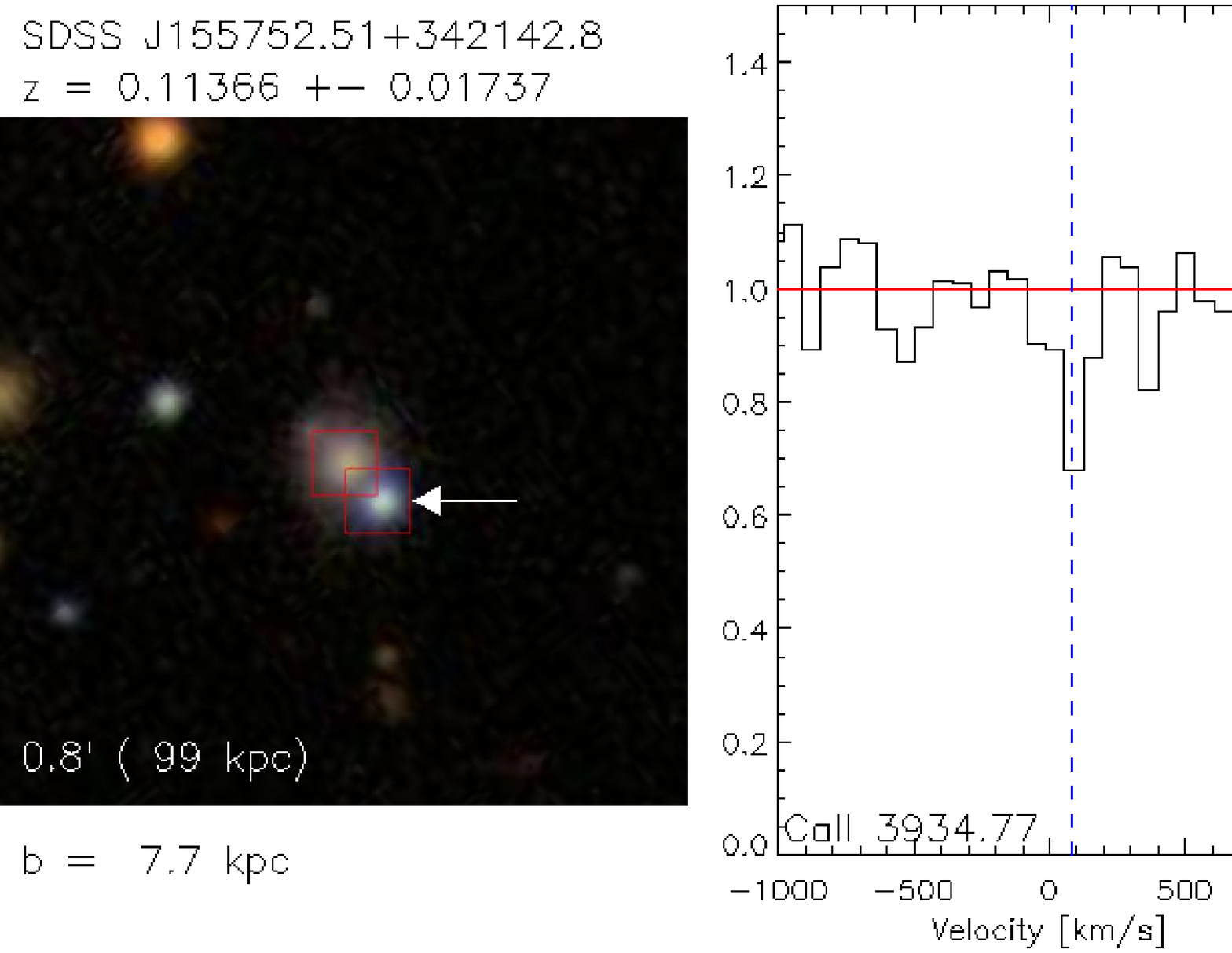}}
\caption{Absorbers with Ca~II and Na~I strong lines.  The figure displays the same format as in Figure~\ref{fig:caiiabs}.}
\label{fig:bothabs}
\end{figure}

\begin{figure}
\includegraphics[scale=0.9]{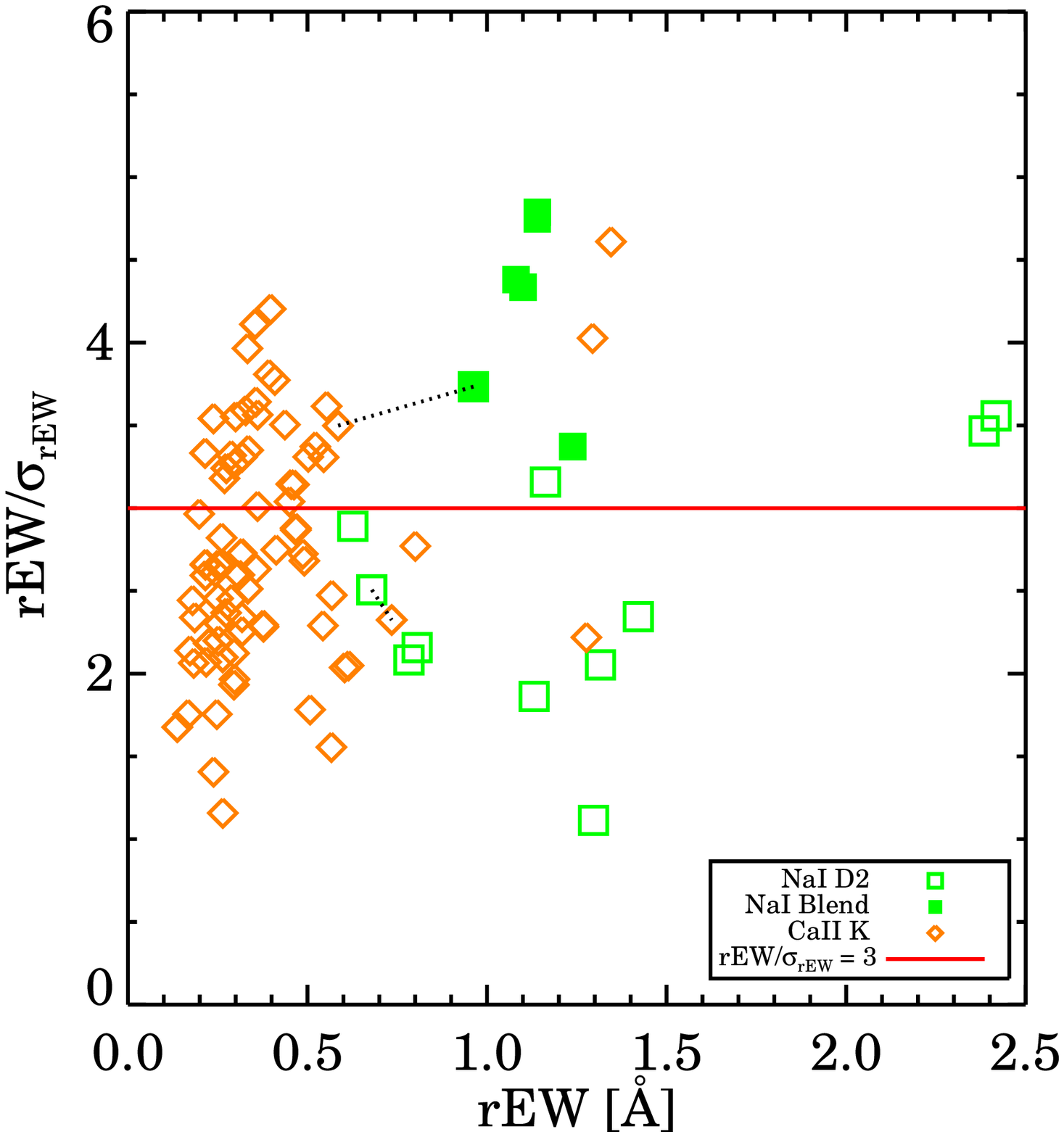}
\caption{Plot of the significance (rEW/$\sigma_{rEW}$) of the Ca~II~K or Na~I~D2 lines vs the rest equivalent width, for the list of absorbers.  Open green squares refer to the Na~I~D2 line.  Filled green squares refer to blended Na~I systems.  Open orange diamonds refer to the Ca~II~K line.  Symbols connected by a dotted line indicate absorbers with both Na~I and Ca~II detected.}
\label{fig:rewvssig}
\end{figure}

\begin{figure}
\includegraphics[scale=0.9]{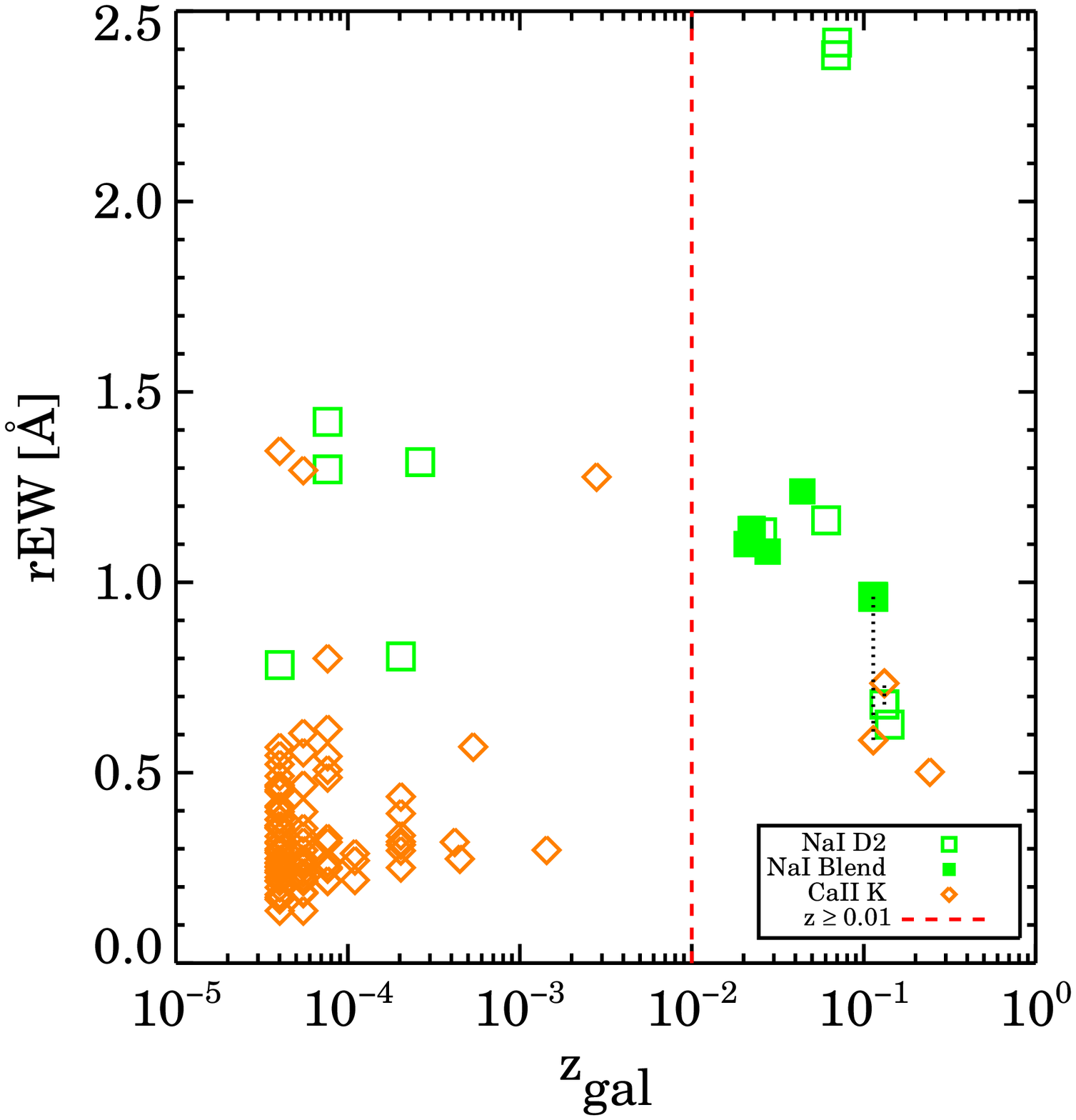}
\caption{Plot of absorber rest equivalent width vs SDSS galaxy redshift, for the list of absorbers. Symbols are the same as in Figure~\ref{fig:rewvssig}.  The dashed vertical line separates absorbers with z$\ge$0.01.  This redshift separator was chosen as a guide to separate absorbers due to Galactic sightlines and Virgo cluster galaxies from `true' extragalactic systems, based on the outer boundary to the Virgo cluster \citep{bing93}.}
\label{fig:rewvszgal}
\end{figure}

\begin{figure}
\includegraphics[scale=0.9]{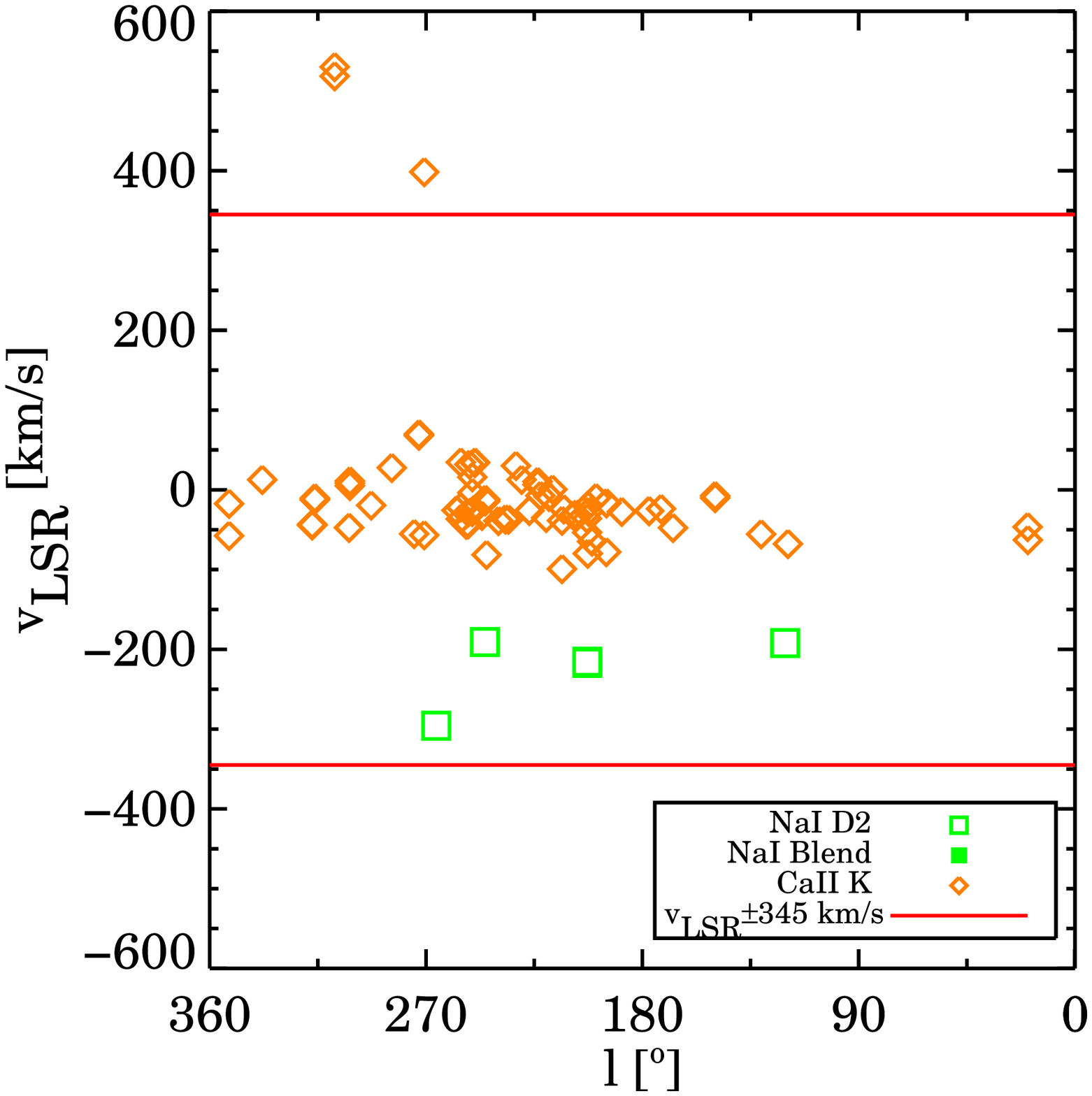}
\caption{Plot of the LSR velocity of the Ca~II~K or Na~I~D2 line (or Na~I doublet, if blended) against the Galactic latitude along the quasar sightline, for the list of absorbers.  The two solid lines at $\pm$345~km~s$^{-1}$ indicate the velocity cutoff separating Galactic High-Velocity Cloud complexes from extragalactic absorbers \citep{wakker91}.  There are several pairs with $v_{LSR}$ $>$ 600~km~s$^{-1}$ not shown on the plot. }
\label{fig:lsrvsl}
\end{figure}

\begin{figure}
\includegraphics[scale=0.9]{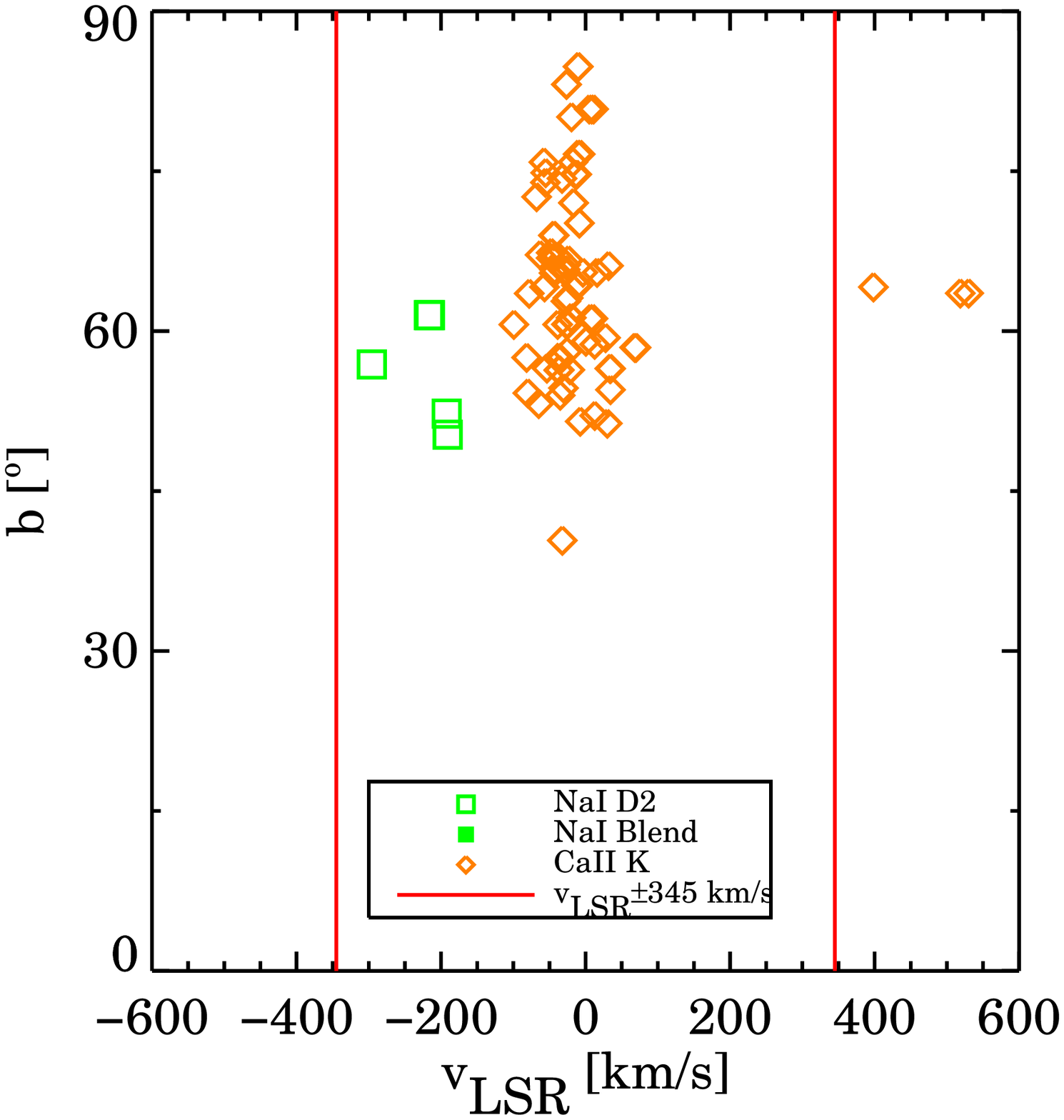}
\caption{Plot of the Galactic latitude against the LSR velocity of the Ca~II~K or Na~I~D2 line (or Na~I doublet, if blended) along the quasar sightline, for the list of absorbers.  The two solid lines at $\pm$345~km~s$^{-1}$ indicate the velocity cutoff separating Galactic High-Velocity Cloud complexes from extragalactic absorbers \citep{wakker91}.  There are several pairs with $v_{LSR}$ $>$ 600~km~s$^{-1}$ not shown on the plot. }
\label{fig:lsrvsb}
\end{figure}

\begin{figure}
\includegraphics[scale=0.9]{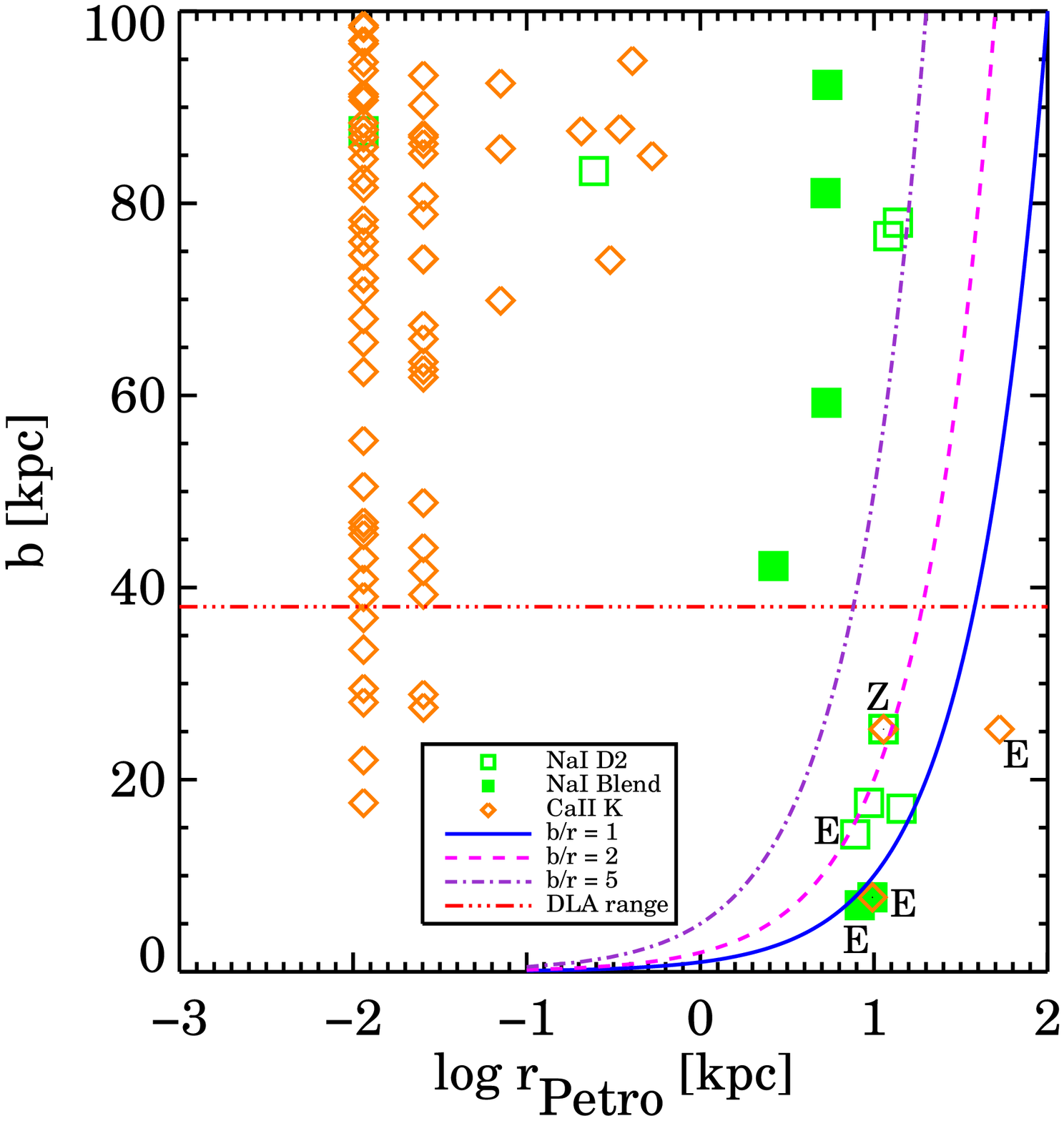}
\caption{Plot of quasar impact parameter versus SDSS r-band Petrosian radius, for the list of absorbers.  The line at 38~kpc represents the impact parameter out to which low-z DLAs have been found by \cite{rao03}.  The three curved lines indicate the ratio of the impact parameter to the r-band Petrosian radius (1, 2, and 5, respectively).  The vertical striping apparent in the data at low Petrosian radii is a result of multiple QSO sightlines being found within the 100~kpc search radius of a low redshift galaxy.  The absorbers marked with `E' and `Z' represent the 4 extragalactic absorbers, and the Zych absorber, discussed in Section 5.}
\label{fig:bkpcvsrpet}
\end{figure}

\begin{figure}
 \includegraphics[scale=0.9]{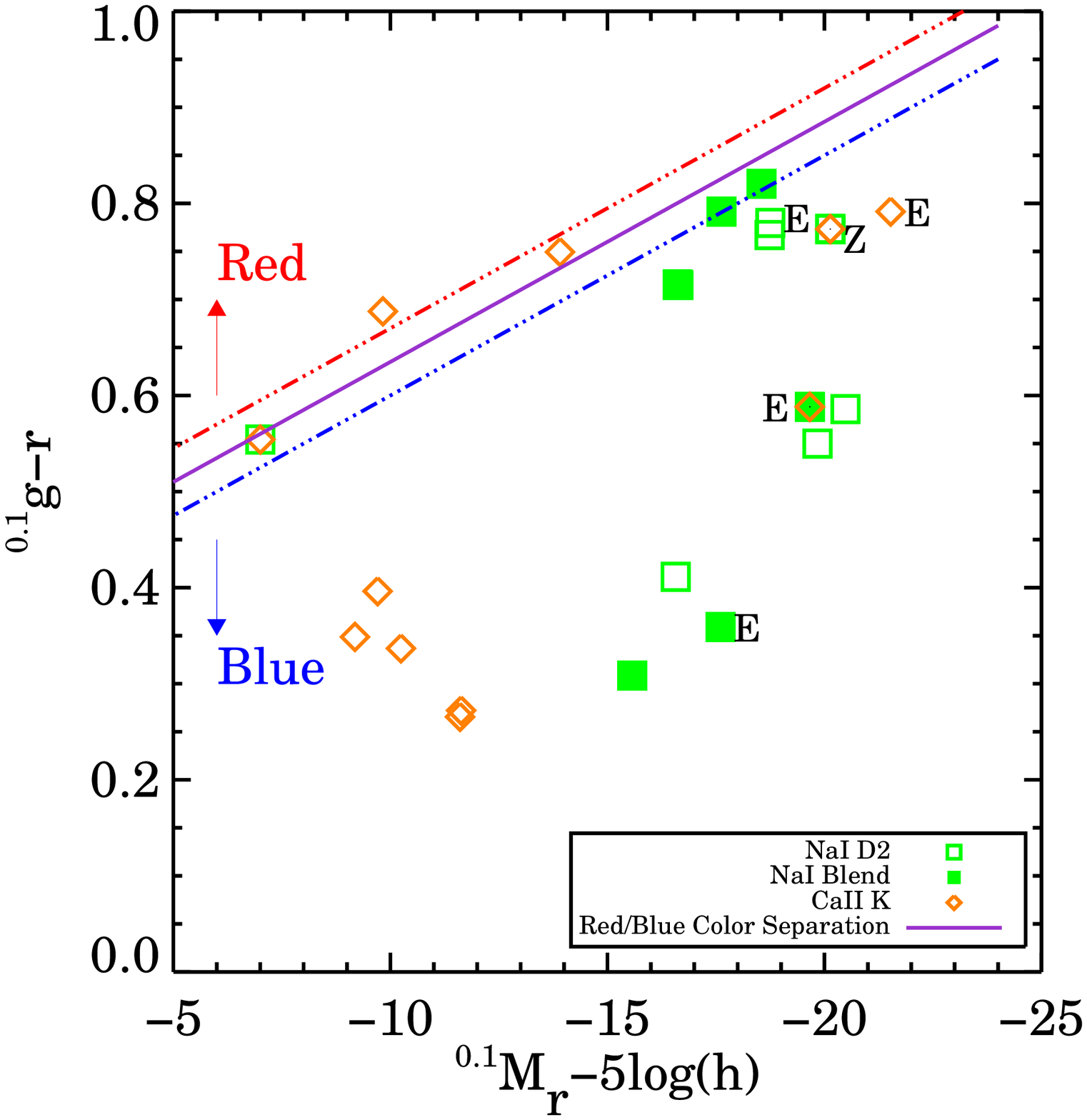}
\caption{Plot of galaxy g-r color versus galaxy SDSS r-band absolute magnitude, for the list of absorbers.  The purple line corresponds to the average of the red/blue galaxy cutoffs from \cite{yan06}.  Applying the cut divides our sample into 5 red galaxies and 19 blue galaxies.  Galaxies between the red and blue regions are more ambiguous in their classification.  The absorbers marked with `E' and `Z' represent the 4 extragalactic absorbers, and the Zych absorber, discussed in Section 5.  All absolute magnitudes were calculated using distances derived from heliocentric redshifts, with the exception of the four Virgo cluster galaxies.  For those galaxies, distances were derived from redshifts corrected for the Local Group infall towards Virgo, taken from HyperLeda \citep{paturel03}.}
\label{fig:grMr}
\end{figure}

%-------------------------------------------------------------------------------------------------------
\clearpage
\begin{landscape}
\begin{table}[tbp]
{\scriptsize
\caption{Parent Sample of 97468 Galaxy-Quasar Pairs}
\label{tab:parent}
\begin{tabular}{lclcccccccccc}
\hline
Galaxy & $z_{sdss}$ & Quasar & $z_{QSO}$ & 100kpc Scale & b & b & inCa & inNa & CaFlag & NaFlag & $VC_{CaII}$ & $VC_{NaI}$ \\
 & & & & ['] & ['] & [kpc] & & & & & & \\
\hline
SDSS J000005.53-011258.9 & 0.03170$\pm$0.00009 & SDSS J235958.66-011225.3 & 1.77180 &    2.63 &    1.80 & 68.63 & 1 & 1 & N & N & - &  - \\
SDSS J000006.67+003016.7 & 0.10916$\pm$0.00015 & SDSS J000006.53+003055.2 & 1.82460 &    0.84 &    0.64 & 76.84 & 1 & 1 & N & N & - &  - \\
SDSS J000027.97-002653.4 & 0.07897$\pm$0.00015 & SDSS J000030.37-002732.4 & 1.80550 &    1.12 &    0.88 & 79.24 & 1 & 1 & N & N & - &  - \\
SDSS J000031.98-103322.0 & 0.07713$\pm$0.00015 & SDSS J000035.75-103305.3 & 1.21820 &    1.14 &    0.97 & 84.75 & 1 & 1 & N & N & - &  - \\
SDSS J024910.78-074924.6 & 0.00457$\pm$0.00001 & SDSS J024945.00-075626.9 & 1.58460 &   17.69 &   11.02 & 62.28 & 1 & 1 & Q & N & 4 &  - \\
SDSS J024910.78-074924.6 & 0.00457$\pm$0.00001 & SDSS J024950.68-075949.9 & 0.97360 &   17.69 &   14.36 & 81.17 & 1 & 1 & N & Q & - & 7 \\
SDSS J024910.78-074924.6 & 0.00457$\pm$0.00001 & SDSS J024957.31-075617.0 & 2.16360 &   17.69 &   13.42 & 75.84 & 1 & 1 & N & N & - &  - \\
SDSS J025346.70-072344.0 & 0.00449$\pm$0.00001 & SDSS J025407.68-074023.0 & 0.73160 &   18.00 &   17.44 & 96.88 & 1 & 1 & N & N & - &  - \\
SDSS J025346.70-072344.0 & 0.00449$\pm$0.00001 & SDSS J025414.61-073435.4 & 1.47220 &   18.00 &   12.87 & 71.51 & 1 & 1 & N & Q & - & 4 \\
SDSS J025346.70-072344.0 & 0.00449$\pm$0.00001 & SDSS J025420.59-073745.5 & 2.16710 &   18.00 &   16.35 & 90.80 & 1 & 1 & Q & N & 3 &  - \\
SDSS J122037.63+283803.3 & 0.02764$\pm$0.00011 & SDSS J122037.22+283752.0 & 2.20420 &    3.00 &    0.21 &  6.92 & 1 & 1 & N & A & - & 1 \\
SDSS J141745.62+162509.3 & 0.24229$\pm$0.00015 & SDSS J141746.03+162512.2 & 1.71690 &    0.44 &    0.11 & 25.26 & 1 & 1 & A & N & 1 &  - \\
\hline
\tablecomments{The columns are 1-SDSS galaxy name, 2-SDSS galaxy redshift, 3-SDSS quasar name, 4-SDSS quasar redshift, 5-The 100~kpc search radius scale in arcmin, 6-impact parameter in arcmin, 7-impact parameter in kpc, 8$\&$9-flags indicating the inclusion (1) or exclusion (0) due to Ly-$\alpha$ forest contamination, 10$\&$11-flags indicating absorber candidate (A), questionable candidate (Q), non-detection (N), interloping doublet (D), removal due to sky line (S), 12$\&$13-flags indicating our visual classification of the absorber or questionable candidate (see the FITS header for a legend).}
\end{tabular}
}
\end{table}
\end{landscape}

% table of individual CaII absorbers
\clearpage
\begin{table}[tbp]
{\small
\caption{Individual Ca~II Absorbers}
\label{tab:uniqCaII}
\begin{tabular}{lccl}
\hline
 Galaxy & $z_{sdss}$ & QSO Count  & Notes\\
\hline
SDSS J091338.99+193707.4 & 0.00143$\pm$0.00001 &  1  &  \\
SDSS J102703.86+283721.9 & 0.00020$\pm$0.00016 &  7  & inconsistent spectrum, not a part of NGC~3245A\\
SDSS J111849.76-002109.9 & 0.13179$\pm$0.00016 &  1  &  see also \cite{zych07}\\
SDSS J113420.50-033525.4 & 0.00008$\pm$0.00007 & 12  & inconsistent spectrum \\
SDSS J114637.59+405036.6 & 0.00280$\pm$0.00002 &  1  &  \\
SDSS J121323.27+295518.4 & 0.00042$\pm$0.00001 &  1  & KUG~1210+301B \\
SDSS J121344.76+363802.4 & 0.00045$\pm$0.00018 &  1 & in NGC~4190 \\
SDSS J121633.70+130153.6 & 0.00006$\pm$0.00007 & 21  & VCC~200 (dE2) \\
SDSS J122843.30+114518.1 & 0.00054$\pm$0.00007 &  1  & VCC~1125 (S0(9)) \\
SDSS J122844.91+124835.1 & 0.00004$\pm$0.00033 & 41  & VCC~1129 (dE3) \\
SDSS J123745.22+070618.3 & 0.00011$\pm$0.00032 &  3  & VCC~1726 (SdmIV) \\
SDSS J141745.62+162509.3 & 0.24229$\pm$0.00015 &  1  &  \\
SDSS J155752.51+342142.8 & 0.11366$\pm$0.00006 &  1  &  \\
\hline
\tablecomments{Inconsistent spectrum refers to an inconsistency between the reported SDSS redshift with the galaxy spectrum.}
\end{tabular}
}
\end{table}

% table of individual NaI absorbers
%\clearpage
\begin{table}[tbp]
{\small
\caption{Individual Na~I Absorbers}
\label{tab:uniqNaI}
\begin{tabular}{lccl}
\hline
 Galaxy & $z_{sdss}$ & QSO Count  & Notes\\
\hline
%SDSS J082555.52+353231.9 & 0.00269$\pm$0.00118 &  1  & blue compact galaxy  \\
%SDSS J083747.62+183946.2 & 0.00196$\pm$0.00016 &  1  & inconsistent spectrum  \\
%SDSS J084840.21+010219.6 & 0.02881$\pm$0.00016 &  1  & UGC~4613   \\
%SDSS J092317.04+515822.6 & 0.00152$\pm$0.00016 &  1  & MCG~+09-16-010  \\
SDSS J102703.86+283721.9 & 0.00020$\pm$0.00016 & 1  & inconsistent spectrum, not in NGC~3245A  \\
%SDSS J103617.59+362531.1 & 0.00204$\pm$0.00001 &  1  & KUG~1033+366B  \\
%SDSS J111425.16+153201.8 & 0.00190$\pm$0.00020 &  1  &   \\
SDSS J111849.76-002109.9 & 0.13179$\pm$0.00016 &  1  &   \\
SDSS J113420.50-033525.4 & 0.00008$\pm$0.00007 & 2  & inconsistent spectrum \\
SDSS J114313.05+193646.9 & 0.02085$\pm$0.00009 &  1  & in Abell 1367  \\
SDSS J114318.07+193401.3 & 0.02262$\pm$0.00015 &  1  & in Abell 1367  \\
SDSS J114336.98+193616.7 & 0.02201$\pm$0.00019 &  1  & in Abell 1367  \\
SDSS J115115.25+485331.0 & 0.02565$\pm$0.00011 &  1  &   \\
%SDSS J115932.83+224147.6 & 0.00152$\pm$0.00016 &  1  & z not ok, new z$\sim$0.408  \\
%SDSS J120526.43+224337.5 & 0.00184$\pm$0.00013 &  1  & z not ok, new z$\sim$0.0757  \\
%SDSS J120847.78+511146.8 & 0.00194$\pm$0.00017 &  1  &   \\
%SDSS J121323.27+295518.4 & 0.00042$\pm$0.00001 &  1  & KUG~1210+301B  \\
%SDSS J121344.76+363802.4 & 0.00045$\pm$0.00018 &  2  & in NGC~4190  \\
%SDSS J121633.70+130153.6 & 0.00006$\pm$0.00007 & 3  & VCC~200 (dE2)  \\
SDSS J122037.63+283803.3 & 0.02764$\pm$0.00011 &  1  &   \\
%SDSS J122130.67+123915.1 & 0.00113$\pm$0.00013 &  1  &   \\
%SDSS J122540.46+333341.9 & 0.00115$\pm$0.00001 &  1  & in NGC~4395   \\
%SDSS J122543.22+333055.1 & 0.00098$\pm$0.00001 &  1  & in NGC~4395   \\
%SDSS J122544.49+333512.3 & 0.00117$\pm$0.00001 &  1  & in NGC~ 4395   \\
%SDSS J122548.86+333248.7 & 0.00104$\pm$0.00094 &  1  & in NGC~4395, center  \\
%SDSS J122553.97+333222.5 & 0.00101$\pm$0.00001 &  1  & in NGC~4395   \\
%SDSS J122554.64+333046.9 & 0.00097$\pm$0.00013 &  1  & in NGC~4395  \\
%SDSS J122559.00+333112.3 & 0.00093$\pm$0.00098 &  1  & in NGC~4395, SA(s)m Sy1.8  \\
%SDSS J122843.30+114518.1 & 0.00054$\pm$0.00007 &  1  & VCC~1125 (SO(9))  \\
SDSS J122844.91+124835.1 & 0.00004$\pm$0.00033 & 1  & VCC~1129 (dE3)  \\
%SDSS J123745.22+070618.3 & 0.00011$\pm$0.00032 & 1  & VCC~1726 (SdmIV)  \\
SDSS J124006.10+613609.5 & 0.00026$\pm$0.00017 &  1  & in NGC~4605  \\
%SDSS J124125.41+191317.5 & 0.00160$\pm$0.00007 &  1  &  \\
%SDSS J125440.95+635325.1 & 0.00192$\pm$0.00004 &  1  &   \\
%SDSS J140253.21+543125.3 & 0.00095$\pm$0.00002 &  1  & in M101  \\
SDSS J140613.22+153035.5 & 0.06040$\pm$0.00018 &  1  &   \\
SDSS J142009.64+132626.7 & 0.14152$\pm$0.00020 &  1  &   \\
%SDSS J145631.01+593540.7 & 0.00155$\pm$0.00013 &  1  & MCG~+10-21-039  \\
SDSS J150400.94+240437.1 & 0.06894$\pm$0.00010 &  1  & in group \\
SDSS J150403.17+240559.8 & 0.06997$\pm$0.00019 &  1  & in group  \\
SDSS J151145.63+071510.6 & 0.04393$\pm$0.00019 &  1  &   \\
\hline
\end{tabular}
}
\end{table}

% table of individual both absorbers (CaII and NaI)
%\clearpage
\begin{table}[tbp]
{\small
\caption{Individual Na~I $\&$ Ca~II Absorbers}
\label{tab:uniqboth}
\begin{tabular}{lccl}
\hline
 Galaxy & $z_{sdss}$ & QSO Count  & Notes\\
\hline
%SDSS J102703.86+283721.9 & 0.00020$\pm$0.00016 &  5  & inconsistent spectrum, not in NGC~3245A \\
SDSS J111849.76-002109.9 & 0.13179$\pm$0.00016 &  1  &  \\
%SDSS J113420.50-033525.4 & 0.00008$\pm$0.00007 &  7  & inconsistent spectrum \\
%SDSS J121344.76+363802.4 & 0.00045$\pm$0.00018 &  1  & in NGC~4190 \\
%SDSS J121633.70+130153.6 & 0.00006$\pm$0.00007 & 1 & VCC~200 (dE2) \\
%SDSS J121843.95+122308.9 & 0.00052$\pm$0.00028 &  1  & VCC~304 (dE1pec) \\
%SDSS J122155.58+115800.5 & 0.00047$\pm$0.00001 &  1  & VCC~512 (Im) \\
%SDSS J122603.87+130643.6 & 0.00045$\pm$0.00014 &  1  & VCC~873 (Sb edge-on) \\
%SDSS J122844.91+124835.1 & 0.00004$\pm$0.00033 & 29  & VCC~1129 (dE3) \\
%SDSS J123745.22+070618.3 & 0.00011$\pm$0.00032 &  5  & VCC~1726 (SdmIV) \\
%SDSS J140253.21+543125.3 & 0.00095$\pm$0.00002 &  1  & in M101 \\
SDSS J155752.51+342142.8 & 0.11366$\pm$0.00006 &  1  &  \\
\hline
\end{tabular}
}
\end{table}

%Table 1 ---------------------------------------------
\clearpage
\begin{landscape}
\begin{table}[tbp]
{\small
\caption{Total Absorber Sample: Position Information}
\label{tab:posinfo}
\begin{tabular}{lccclcccccl}
\tableline
Galaxy	&	$z_G$	&	{\it l}	&	{\it b}	&	QSO	&	$z_{QSO}$	&	{\it l}	&	{\it b}	&	b	&	b/$r_{Petro}$	&	Ion \\
	&		&	[$^{\circ}$]	&	[$^{\circ}$]	&		&		&	[$^{\circ}$]	&	[$^{\circ}$]	&	[kpc]	&		&	 \\
\tableline\tableline
 SDSS J091338.99+193707.4 & 0.00143$\pm$0.00001 & 209 &  40 & SDSS J091511.03+201248.3 & 1.23900 & 209 &  40 & 74.13 &   245.1 & CaII \\
\\[1pt]
SDSS J102703.86+283721.9 & 0.00020$\pm$0.00016 & 202 &  58 & SDSS J100417.96+282444.1 & 0.32820 & 201 &  53 & 75.87 &   411.5 & CaII \\
 & 0.00020$\pm$0.00016 & 202 &  58 & SDSS J100927.44+273215.7 & 1.51230 & 203 &  54 & 61.12 &   331.6 & CaII \\
 & 0.00020$\pm$0.00016 & 202 &  58 & SDSS J101353.43+244916.4 & 1.63430 & 208 &  55 & 72.86 &   395.2 & CaII \\
 & 0.00020$\pm$0.00016 & 202 &  58 & SDSS J101956.59+274401.7 & 1.92500 & 203 &  57 & 27.34 &   148.3 & CaII \\
 & 0.00020$\pm$0.00016 & 202 &  58 & SDSS J104111.97+282805.0 & 0.21100 & 203 &  61 & 47.11 &   255.6 & CaII \\
 & 0.00020$\pm$0.00016 & 202 &  58 & SDSS J104221.97+282013.3 & 1.68590 & 203 &  61 & 51.15 &   277.5 & NaI \\
 & 0.00020$\pm$0.00016 & 202 &  58 & SDSS J104224.85+231001.7 & 1.23720 & 213 &  61 & 97.85 &   530.8 & CaII \\
 & 0.00020$\pm$0.00016 & 202 &  58 & SDSS J105124.28+320044.5 & 0.44600 & 195 &  64 & 94.74 &   513.9 & CaII \\
\\[1pt]
SDSS J111849.76-002109.9 & 0.13179$\pm$0.00016 & 260 &  55 & SDSS J111850.13-002100.7 & 1.02560 & 260 &  55 & 25.25 &     2.2 & NaI \\
 & 0.13179$\pm$0.00016 & 260 &  55 & SDSS J111850.13-002100.7 & 1.02560 & 260 &  55 & 25.25 &     2.2 & CaII \\
 & 0.13179$\pm$0.00007 & 260 &  55 & SDSS J111850.13-002100.7 & 1.02560 & 260 &  55 & 25.25 &     2.2 & Both \\
\\[1pt]
SDSS J113420.50-033525.4 & 0.00008$\pm$0.00007 & 269 &  54 & SDSS J104102.43+023242.8 & 1.95870 & 246 &  50 & 83.38 &  9958.3 & NaI \\
 & 0.00008$\pm$0.00007 & 269 &  54 & SDSS J110156.34+073525.2 & 1.51360 & 245 &  58 & 78.45 &  9369.3 & CaII \\
 & 0.00008$\pm$0.00007 & 269 &  54 & SDSS J110557.87+045728.3 & 0.90850 & 250 &  56 & 63.14 &  7541.0 & CaII \\
 & 0.00008$\pm$0.00007 & 269 &  54 & SDSS J111012.07+011327.8 & 0.09500 & 256 &  55 & 43.89 &  5241.4 & CaII \\
 & 0.00008$\pm$0.00007 & 269 &  54 & SDSS J111507.65+023757.5 & 0.56650 & 256 &  56 & 44.69 &  5337.7 & CaII \\
 & 0.00008$\pm$0.00007 & 269 &  54 & SDSS J111816.94+074558.2 & 1.73500 & 250 &  61 & 68.47 &  8178.1 & CaII \\ 
\tableline
\tablecomments{Ion refers to which line list the absorber belongs to.  Absorbers listed as Na~I have the Na~I doublet detected.  Absorbers listed as Ca~II have the Ca~II doublet detected.  Absorbers listed as Both have at least both the Ca~II~K and Na~I~D2 lines detected.  The full table is provided with the online version of the paper.}
\end{tabular}
\label{tab:samppos}
}
\end{table}
\end{landscape}

%Table 2 ------------------------------------------------------------------------
\clearpage
\begin{landscape}
\begin{table}[tbp]
{\scriptsize
\caption{Total Absorber Sample: Line Measurements}
\label{tab:lineinfo}
\begin{tabular}{lclcccccccccccccc}
\tableline
Galaxy	&	$z_G$	&	Ion	&	$v_{LSR}$	&	$rEW_1$	&	$rEW_1$/$\sigma_{rEW_1}$	&	$z_{line1}$	&	$\Delta v_1$	&	$rEW_2$	&	$z_{line2}$	&	$\Delta v_2$ \\
	&		&		&	[km s$^{-1}$]	&	[\AA]	&		&		&	[km s$^{-1}$]	&	[\AA]	&		&	[km s$^{-1}$] \\
\tableline\tableline
 SDSS J091338.99+193707.4 & 0.00143$\pm$0.00001 & CaII &   -32.2 &  0.30$\pm$ 0.15 &  1.97 & -0.00009$\pm$0.00047 & -455.6$\pm$ 139.9 &  0.29$\pm$ 0.13 & -0.00008$\pm$0.00047 &  -453.1$\pm$ 140.1 \\
\\[1pt]
SDSS J102703.86+283721.9 & 0.00020$\pm$0.00016 & CaII &   -64.8 &  0.25$\pm$ 0.09 &  2.66 & -0.00021$\pm$0.00051 & -123.8$\pm$ 160.1 &  0.21$\pm$ 0.12 & -0.00059$\pm$0.00055 &  -237.8$\pm$ 172.4 \\
 & 0.00020$\pm$0.00016 & CaII &   -79.9 &  0.44$\pm$ 0.12 &  3.50 & -0.00026$\pm$0.00038 & -139.0$\pm$ 125.5 &  0.18$\pm$ 0.12 & -0.00015$\pm$0.00059 &  -104.8$\pm$ 183.1 \\
 & 0.00020$\pm$0.00016 & CaII &   -30.6 &  0.39$\pm$ 0.10 &  3.81 & -0.00009$\pm$0.00041 &  -89.4$\pm$ 131.3 &  0.32$\pm$ 0.10 & -0.00024$\pm$0.00044 &  -132.7$\pm$ 141.8 \\
 & 0.00020$\pm$0.00016 & CaII &   -53.5 &  0.31$\pm$ 0.12 &  2.60 & -0.00017$\pm$0.00046 & -113.1$\pm$ 145.5 &  0.23$\pm$ 0.13 &  0.00015$\pm$0.00053 &   -15.8$\pm$ 165.6 \\
 & 0.00020$\pm$0.00016 & CaII &   -19.8 &  0.34$\pm$ 0.10 &  3.35 & -0.00007$\pm$0.00044 &  -80.7$\pm$ 140.7 &  0.29$\pm$ 0.10 & -0.00016$\pm$0.00047 &  -109.8$\pm$ 149.8 \\
 & 0.00020$\pm$0.00016 & NaI &  -216.9 &  0.81$\pm$ 0.37 &  2.16 & -0.00072$\pm$0.00019 & -277.8$\pm$  75.2 &  0.59$\pm$ 0.28 & -0.00046$\pm$0.00022 &  -199.1$\pm$  82.8 \\
 & 0.00020$\pm$0.00016 & CaII &   -38.6 &  0.30$\pm$ 0.15 &  1.93 & -0.00013$\pm$0.00047 &  -98.7$\pm$ 148.8 &  0.23$\pm$ 0.14 &  0.00001$\pm$0.00052 &   -56.8$\pm$ 165.0 \\
 & 0.00020$\pm$0.00016 & CaII &   -77.9 &  0.32$\pm$ 0.14 &  2.26 & -0.00026$\pm$0.00045 & -139.9$\pm$ 143.9 &  0.36$\pm$ 0.15 & -0.00064$\pm$0.00042 &  -251.6$\pm$ 135.5 \\
\\[1pt]
SDSS J111849.76-002109.9 & 0.13179$\pm$0.00016 & NaI & 39538.5 &  0.68$\pm$ 0.27 &  2.51 &  0.13180$\pm$0.00019 &    2.3$\pm$  75.7 &  0.31$\pm$ 0.27 &  0.13187$\pm$0.00029 &    22.1$\pm$  98.7 \\
 & 0.13179$\pm$0.00016 & CaII & 39400.1 &  0.73$\pm$ 0.32 &  2.32 &  0.13134$\pm$0.00028 & -136.1$\pm$  96.7 &  0.62$\pm$ 0.28 &  0.13176$\pm$0.00030 &   -10.6$\pm$ 102.5 \\
 & 0.13179$\pm$0.00007 & Both$_{NaI}$ & 39544.4 &  0.70$\pm$ 0.24 &  2.89 &  0.13182$\pm$0.00019 &    8.2$\pm$  75.1 &  0.31$\pm$ 0.27 &  0.13187$\pm$0.00029 &    22.1$\pm$  98.7 \\
 & 0.13179$\pm$0.00016 & Both$_{CaII}$ & 39544.4 &  0.60$\pm$ 0.23 &  2.58 &  0.13154$\pm$0.00031 &  -76.7$\pm$ 104.8 &  0.63$\pm$ 0.25 &  0.13175$\pm$0.00030 &   -14.3$\pm$ 101.9 \\
\\[1pt]
SDSS J113420.50-033525.4 & 0.00008$\pm$0.00007 & NaI &  -190.8 &  1.30$\pm$ 1.16 &  1.11 & -0.00062$\pm$0.00015 & -209.7$\pm$  49.4 &  0.82$\pm$ 0.77 & -0.00024$\pm$0.00019 &   -94.0$\pm$  59.8 \\
 & 0.00008$\pm$0.00007 & CaII &   -81.5 &  0.51$\pm$ 0.28 &  1.78 & -0.00027$\pm$0.00036 & -102.4$\pm$ 109.0 &  0.37$\pm$ 0.23 & -0.00008$\pm$0.00042 &   -46.0$\pm$ 126.5 \\
 & 0.00008$\pm$0.00007 & CaII &    33.6 &  0.33$\pm$ 0.09 &  3.59 &  0.00012$\pm$0.00044 &   12.8$\pm$ 134.8 &  0.23$\pm$ 0.10 &  0.00019$\pm$0.00053 &    34.0$\pm$ 160.2 \\
 & 0.00008$\pm$0.00007 & CaII &    34.5 &  0.32$\pm$ 0.12 &  2.73 &  0.00012$\pm$0.00045 &   14.0$\pm$ 136.9 &  0.23$\pm$ 0.11 & -0.00019$\pm$0.00052 &   -78.6$\pm$ 157.7 \\
 & 0.00008$\pm$0.00007 & CaII &   -21.8 &  0.49$\pm$ 0.18 &  2.72 & -0.00007$\pm$0.00036 &  -42.8$\pm$ 111.2 &  0.59$\pm$ 0.18 & -0.00011$\pm$0.00033 &   -55.5$\pm$ 100.6 \\
 & 0.00008$\pm$0.00007 & CaII &   -24.8 &  0.22$\pm$ 0.08 &  2.59 & -0.00008$\pm$0.00055 &  -46.7$\pm$ 165.5 &  0.10$\pm$ 0.09 & -0.00010$\pm$0.00081 &   -52.8$\pm$ 243.8 \\
\tableline
\tablecomments{If the Na~I doublet is blended, the values listed for line 1 are measurements based on the blend, while the values for line 2 are blanked out.  The full table is provided in the online version of the paper.}
\label{tab:samplines}
\end{tabular}
}
\end{table}
\end{landscape}

\end{document}